\documentclass[twocolumn]{aastex6}

\newcommand{\ergs}{erg~s$^{-1}$}

\begin{document}
\title{\textit{Hubble Space Telescope} imaging of the active dwarf galaxy RGG 118}

\author{Vivienne F. Baldassare\altaffilmark{1,2}, Amy E. Reines\altaffilmark{3,4}, Elena Gallo\altaffilmark{5}, Jenny E. Greene\altaffilmark{6}}

\altaffiltext{1}{Department of Astronomy, Yale University, New Haven, CT 06511}
\altaffiltext{2}{Einstein Fellow}
\altaffiltext{3}{National Optical Astronomy Observatory, 950 N Cherry Ave, Tucson, AZ 85719}
\altaffiltext{4}{Hubble Fellow}
\altaffiltext{5}{Department of Astronomy, University of Michigan, Ann Arbor, MI 48109}
\altaffiltext{6}{Department of Astrophysical Sciences, Princeton University, Princeton, NJ 08544}


\begin{abstract}
RGG 118 (SDSS 1523+1145) is a nearby ($z=0.0243$), dwarf disk galaxy ($M_{\ast}\approx2\times10^{9} M_{\odot}$) found to host an active $\sim50,000$ solar mass black hole at its core \citep{2015ApJ...809L..14B}. RGG 118 is one of a growing collective sample of dwarf galaxies known to contain active galactic nuclei -- a group which, until recently, contained only a handful of objects. Here, we report on new \textit{Hubble Space Telescope} Wide Field Camera 3 UVIS and IR imaging of RGG 118, with the main goal of analyzing its structure. Using 2-D parametric modeling, we find that the morphology of RGG 118 is best described by an outer spiral disk, inner component consistent with a pseudobulge, and central PSF. The luminosity of the PSF is consistent with the central point source being dominated by the AGN. We measure the luminosity and mass of the ``pseudobulge" and confirm that the central black hole in RGG 118 is under-massive with respect to the $M_{BH}-M_{\rm bulge}$ and $M_{BH}-L_{\rm bulge}$ relations. This result is consistent with a picture in which black holes in disk-dominated galaxies grow primarily through secular processes. 
\end{abstract}

\section{Introduction}

While BHs are ubiquitous in the cores of massive galaxies, the population of BHs in dwarf galaxies ($M_{\ast}<10^{9.5}M_{\odot}$) has been relatively elusive \citep{2016PASA...33...54R}. The first dwarf galaxies identified to have active galactic nuclei (AGNs) were NGC 4395 \citep{1989AJ.....97..726F, 2003ApJ...588L..13F} and Pox 52 \citep{1987AJ.....93...29K}. The AGNs in these systems were serendipitous discoveries, and they were the only dwarf galaxies known to contain AGNs for almost two decades \citep{2004ApJ...607...90B}. In recent years, thanks to large-scale surveys such as the Sloan Digital Sky Survey (SDSS), we have started to identify an increasing number of such systems. Using optical spectroscopic diagnostics, \cite{Reines:2013fj} identified 151 dwarf galaxies with signatures of AGN activity in the SDSS. This constituted an order of magnitude increase in the number of known dwarf galaxies with AGN. While optical spectroscopic diagnostics have identified the largest number of such systems (see also earlier works by \citealt{2004ApJ...610..722G, 2007ApJ...670...92G, 2008AJ....136.1179B}, more recent studies by \citealt{2014AJ....148..136M, 2015MNRAS.454.3722S}), searches using radio and/or X-rays have also been successful at identifying dwarf galaxies with AGNs \citep{:kj,Reines:2011fr, 2014ApJ...787L..30R, 2015ApJ...805...12L, 2016ApJ...831..203P, 2017ApJ...837...48C}. There have been efforts to use IR diagnostics \citep{2014ApJ...784..113S, 2015MNRAS.454.3722S}, though extreme star forming dwarf galaxies can have IR colors that mimic AGNs \citep{2016ApJ...832..119H}. In all, there now exists a collective sample of roughly two hundred dwarf galaxies with AGN signatures. 

With the number of known dwarf galaxies hosting AGN growing, it is important to characterize the host galaxies in detail in order to understand what factors (if any) may influence the presence of an AGN. Additionally, studies of the host galaxies are necessary to determine whether scaling relations between BH mass and host galaxy properties hold at the low-mass end (see \citealt{Kormendy:2013ve} for a review of scaling relations). Where these low-mass systems fall with respect to scaling relations has important implications for BH formation and growth scenarios (\citealt{2008MNRAS.383.1079V, 2012NatCo...3E1304G, 2014GReGr..46.1702N}). For example, semi-analytic models suggest the slope and scatter of the low-mass end of the $M_{\rm BH}-\sigma_{\ast}$ relation between BH mass and bulge stellar velocity dispersion depends on the mechanism by which the first BH seeds formed (\citealt{2009MNRAS.400.1911V}, see also \citealt{:fl, 2016PASA...33...51L} for reviews of BH formation scenarios). 

In the continuing effort towards detailed characterization of host galaxies, we present a \textit{Hubble Space Telescope} imaging analysis of RGG 118 (SDSS 1523+1145), a nearby ($z=0.0243$) dwarf, disk galaxy with an active $\sim50,000~M_{\odot}$ BH. \citep{2015ApJ...809L..14B}. It was first identified as having AGN signatures in \cite{Reines:2013fj} based on narrow emission line ratios which place it in the composite region of the BPT diagram \citep{1981PASP...93....5B, 2003MNRAS.346.1055K, 2006MNRAS.372..961K}. Subsequent analysis of high-resolution spectroscopy with the Magellan Echellette Spectrograph on the 6.5m Clay telescope at Las Campanas Observatory clearly revealed broad H$\alpha$ $\lambda$6563 emission feature characteristic of dense gas orbiting a central massive black hole. Furthermore, the galaxy was found to have a hard X-ray point source coincident with the nucleus -- strong confirmation that RGG 118 hosts an AGN. The mass of the BH, based on single-epoch spectroscopic techniques using the broad H$\alpha$ emission line \citep{2005ApJ...630..122G}, was found to be just $\sim50,000$ solar masses, the smallest yet identified in a galaxy nucleus \citep{2015ApJ...809L..14B}. 

Previous analyses of the morphology of RGG 118 have relied on relatively shallow SDSS imaging \citep{2015ApJ...809L..14B, 2016ApJ...818..172G}. Using 2-D light profile modeling techniques, \cite{2015ApJ...809L..14B} find that RGG 118 is composed of an extended disk, central bulge-like component, and central point source. Subsequent analysis of the SDSS imaging was done by \cite{2016ApJ...818..172G}, who claim the presence of a stellar bar. 
In this paper, we analyze new \textit{Hubble Space Telescope} imaging of RGG 118, with the aim of characterizing the morphology of the host galaxy, and studying the galaxy's stellar populations. 

\section{Data}

We obtained \textit{Hubble Space Telescope} (HST) Wide Field Camera 3 (WFC3) imaging of RGG 118. Images were taken over three orbits during July 2016 (Cycle 23, Proposal 14187, PI: Baldassare). We took observations in two UVIS filters (F475W and F775W) and one IR filter (F160W). These filters correspond to \textit{g}, \textit{i}, and \textit{H} band, respectively. We also employ a traditional four point dither pattern. 

\begin{figure}
\centering
\includegraphics[width=0.44\textwidth]{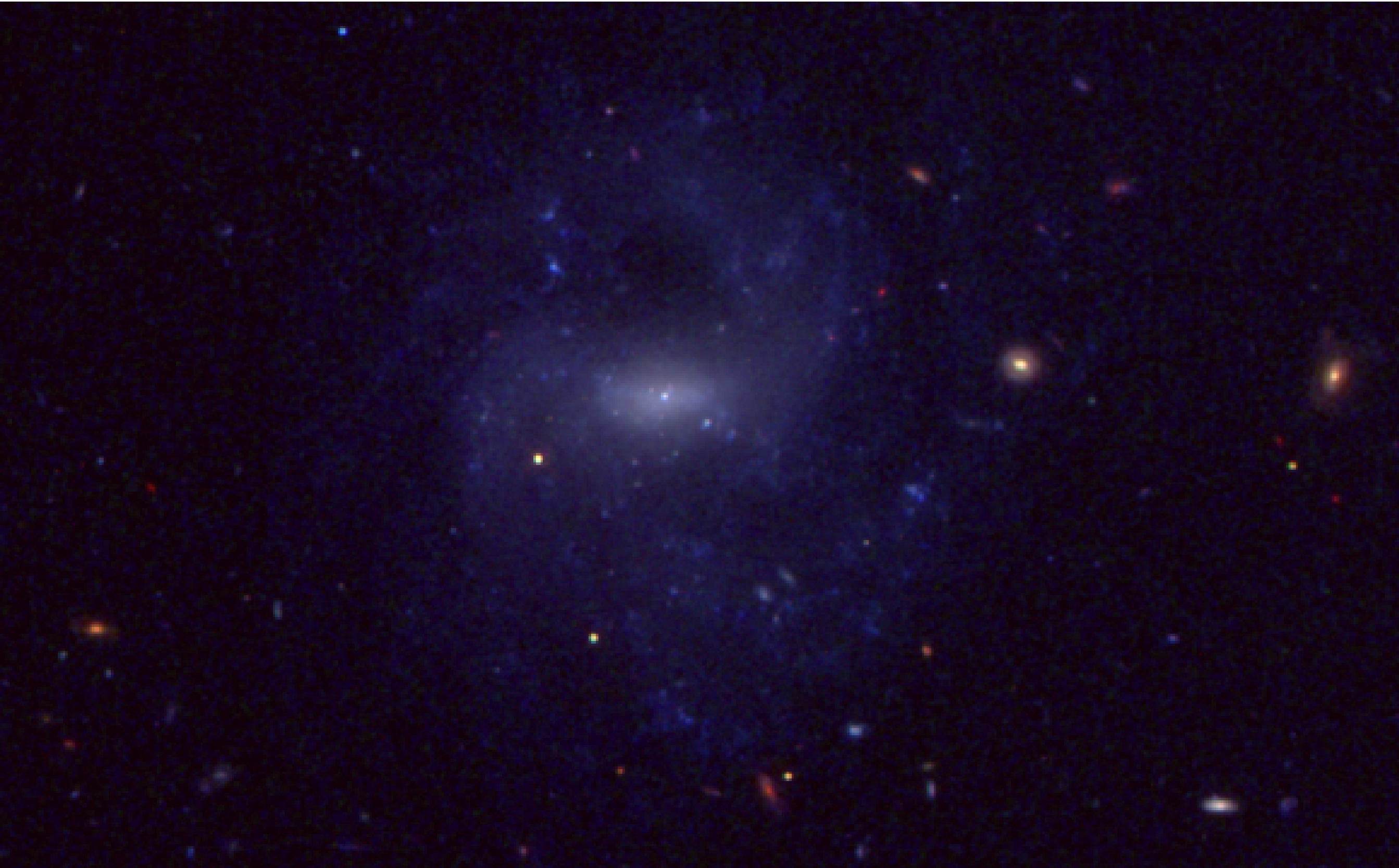}
\caption{Three color image of RGG 118. The UVIS filters have been matched in angular resolution to the IR image (0.09$''$/pixel). The red, green, and blue correspond to the F160W, F775W, and F475W filters, respectively. The images are log-scaled, and the upper and lower scale limits for each filter have been chosen such that the background is black and bright stars are white.}
\label{threecolor}
\end{figure}

\begin{figure*}
\includegraphics[width=0.33\textwidth]{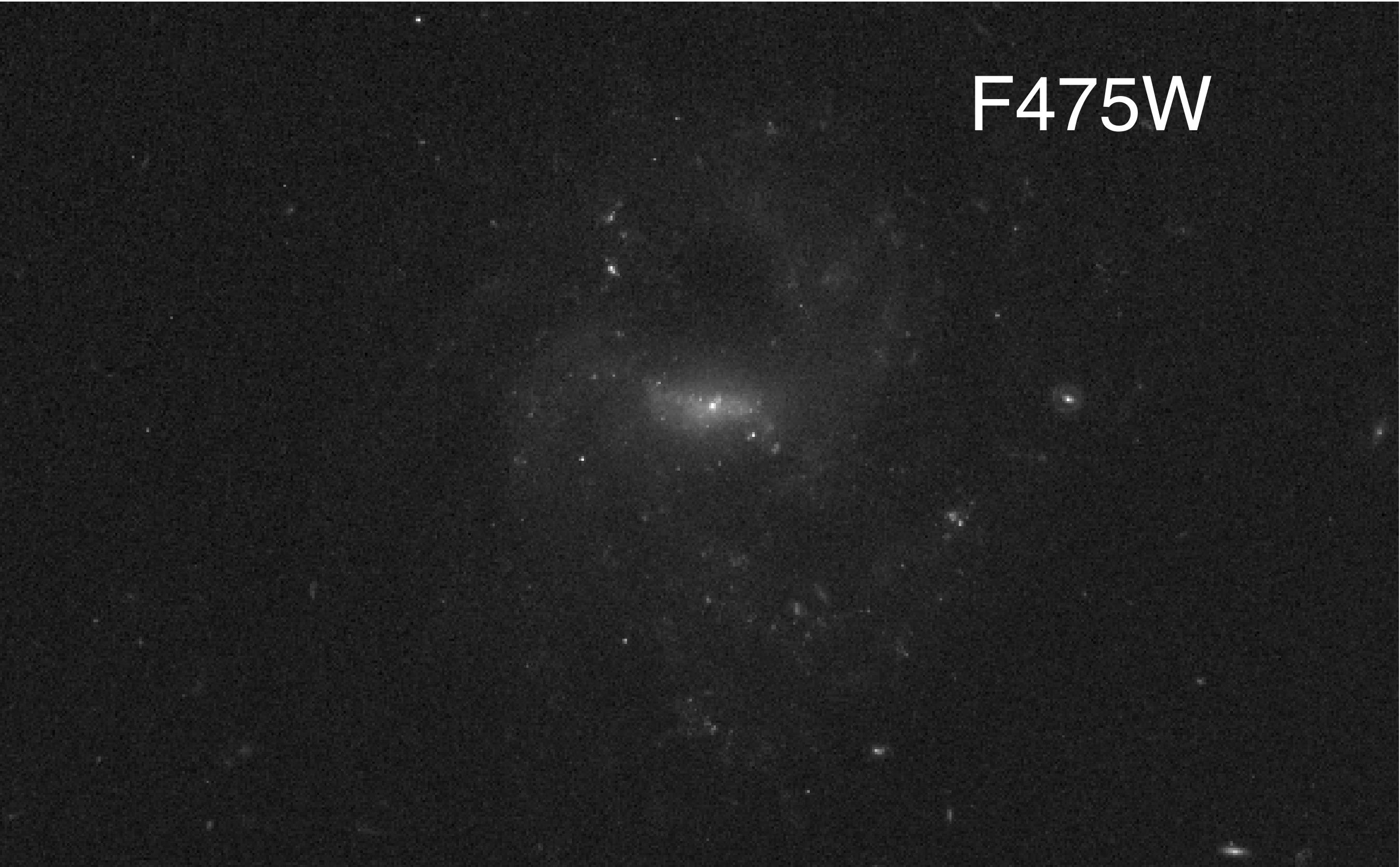}
\includegraphics[width=0.33\textwidth]{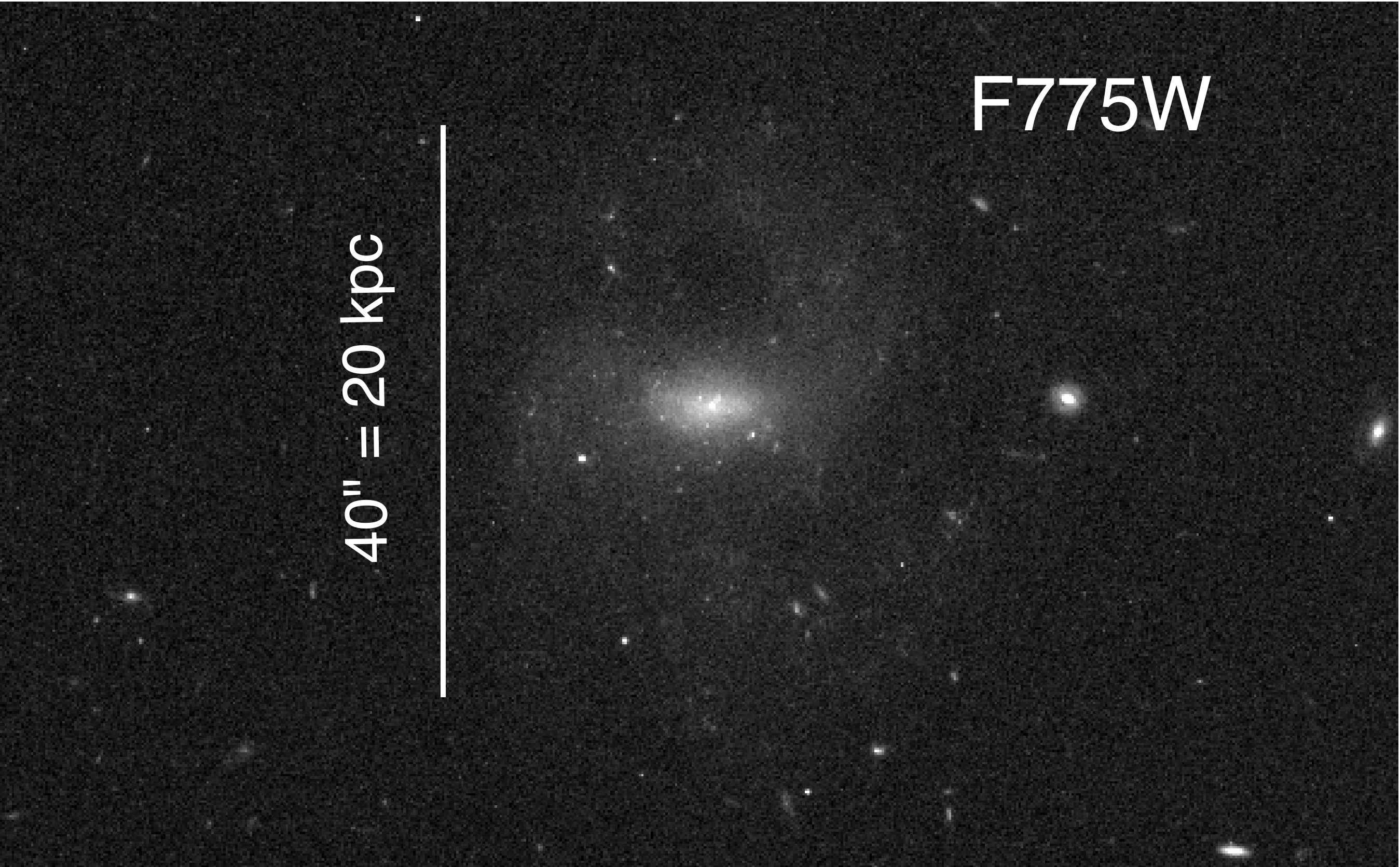}
\includegraphics[width=0.33\textwidth]{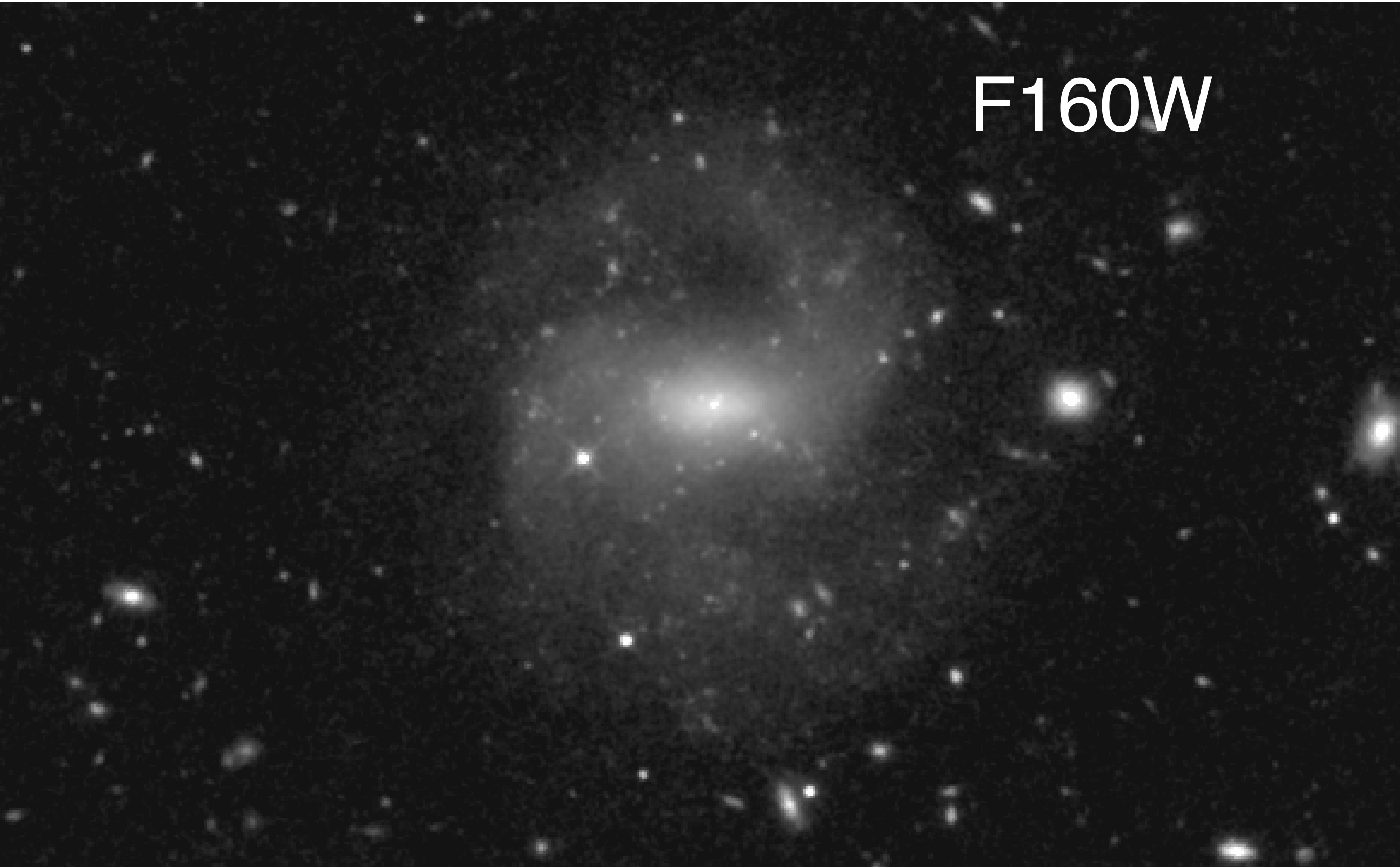} \\
\caption{\textit{Hubble Space Telescope} WFC3 images of RGG 118 in F475W (left), F775W (middle) and F160W (right) filters. Full galaxy images, smoothed with a Gaussian kernel of 3 pixels. The color distribution for all images is in log-scale, and the limits are chosen to encompass the distribution of pixel values.}
\label{final_ims}
\end{figure*}

\begin{figure}
\includegraphics[width=0.5\textwidth]{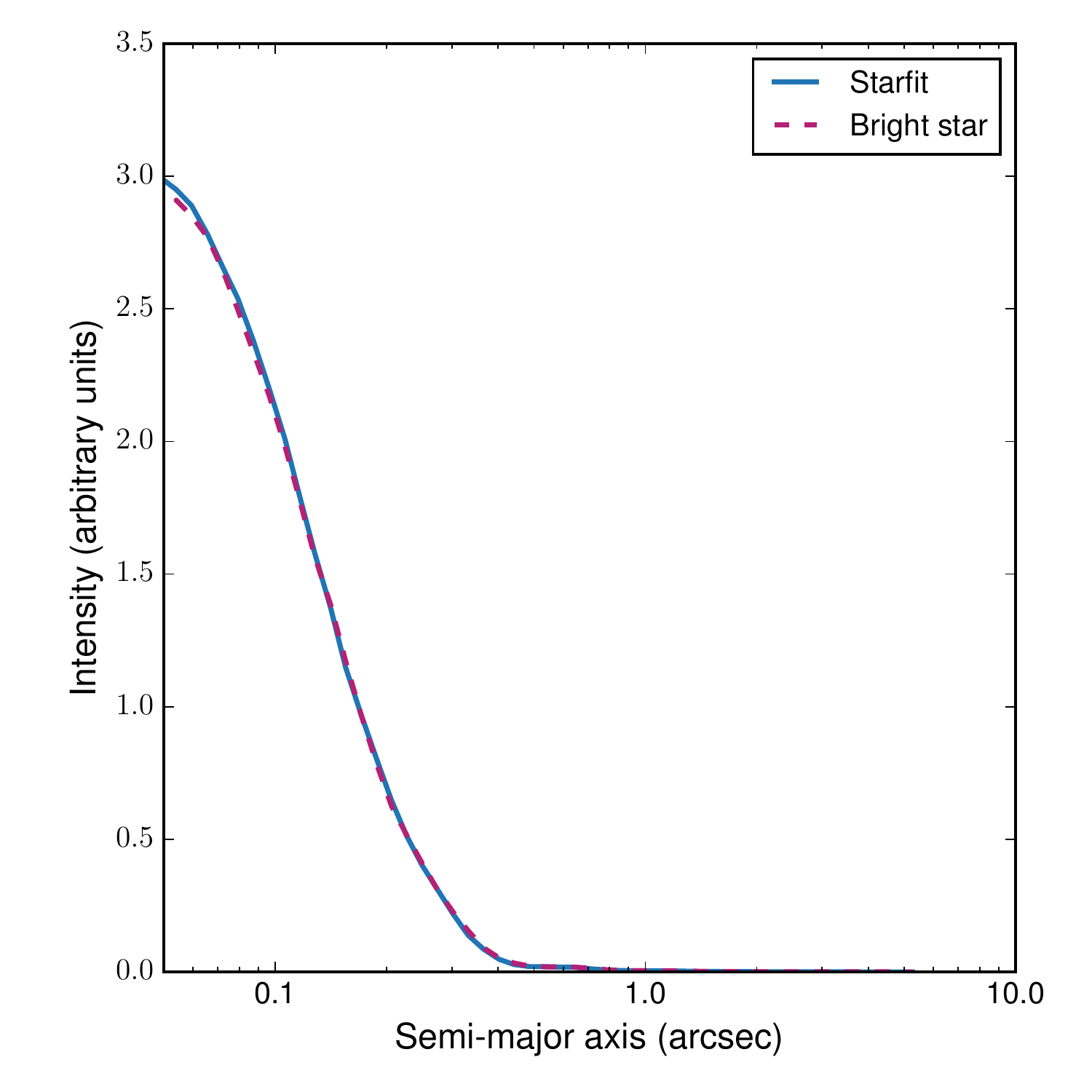}
\caption{Intensity versus semi-major axis of the PSF generated by Starfit and a bright star in the image. They have been normalized to the same central intensity.}
\label{psf_comp}
\end{figure}

Data were reprocessed using the AstroDrizzle pipeline in the DrizzlePac software package. We used a square drizzling kernel and inverse-variance map weighting, recommended for background-limited targets. The native pixel scales for WFC3 are 0.04$''$/pix for the UVIS channel and 0.13$''$/pix for the IR channel. However, dithering of observations allows one to improve the pixel sampling of the final product. For the UVIS observations, our final drizzled product used a final pixel fraction (\textit{final\_pixfrac} parameter in AstroDrizzle) of 0.5 and a final pixel scale (\textit{final\_scale}) of 0.03$''$/pix. The IR observations have a \textit{final\_pixfrac} of 0.8 and \textit{final\_scale} of 0.09$''$/pix. Point spread functions (PSFs) for each filter were constructed using the PSF fitting software \textit{Starfit}\footnote{https://www.ssucet.org/$\sim$thamilton/research/starfit.html}. Figure~\ref{threecolor} shows a three-color \textit{HST} image of RGG 118, and Figure~\ref{final_ims} shows the \textit{HST} imaging in each band. We also construct a PSF using a bright star, in order to determine how much the PSF used impacts the final fit parameters. Figure~\ref{psf_comp} shows a comparison of the Starfit-generated PSF to the profile of a bright star in the F160W image.

\section{Results}
\subsection{Profile Fitting}

We fit the 2-D light profile of RGG 118 using GALFIT \citep{2002AJ....124..266P, 2010AJ....139.2097P}. Fitting is performed on the image taken in the F160W filter, which has the greatest sensitivity. The best fit model is then applied to the optical filters in order to measure the total luminosity of each component. We also compare the results of the 2-D fitting to the 1-D surface brightness profile. We extract 1-D light profiles for each filter using the IRAF program \textit{ellipse}, which fits elliptical isophotes to imaging data. Using the results of \textit{ellipse}, we plot 1-D surface brightness profiles for RGG 118 (i.e., surface brightness as a function of semi-major axis). We also obtain measurements of the ellipticity and position angle as a function of semi-major axis from \textit{ellipse}.

The main goal of this analysis is to decompose the 2-D light profile of RGG 118 into its individual components. Each tested model is comprised of some combination of the following components: S{\'e}rsic profile \citep{1963BAAA....6...41S}, disk (defined as a S{\'e}rsic profile with index $n=1$), Ferrers profile (typically used to model galaxy bars; \citealt{2010AJ....139.2097P}), and PSF. In some models, we also introduce spiral structure in the outermost component. The components used for each tested model are listed in Table~\ref{models}. 

We start by testing a model with a single S{\'e}rsic component to describe the galaxy light output, and find that a single S{\'e}rsic profile produces a poor fit. The addition of a central PSF components improves the fit, but still results in large residuals. We next consider models with two main components: an ``inner" component and an ``outer" component. The outer component is always described by a S{\'e}rsic profile, the index of which is either free to vary or restricted to the canonical disk value of $n=1$. We also consider spiral structure in the outer profile. The inner component is either modeled with a S{\'e}rsic or modified Ferrers profile. 

\begin{figure*}
\includegraphics[width=0.33\textwidth]{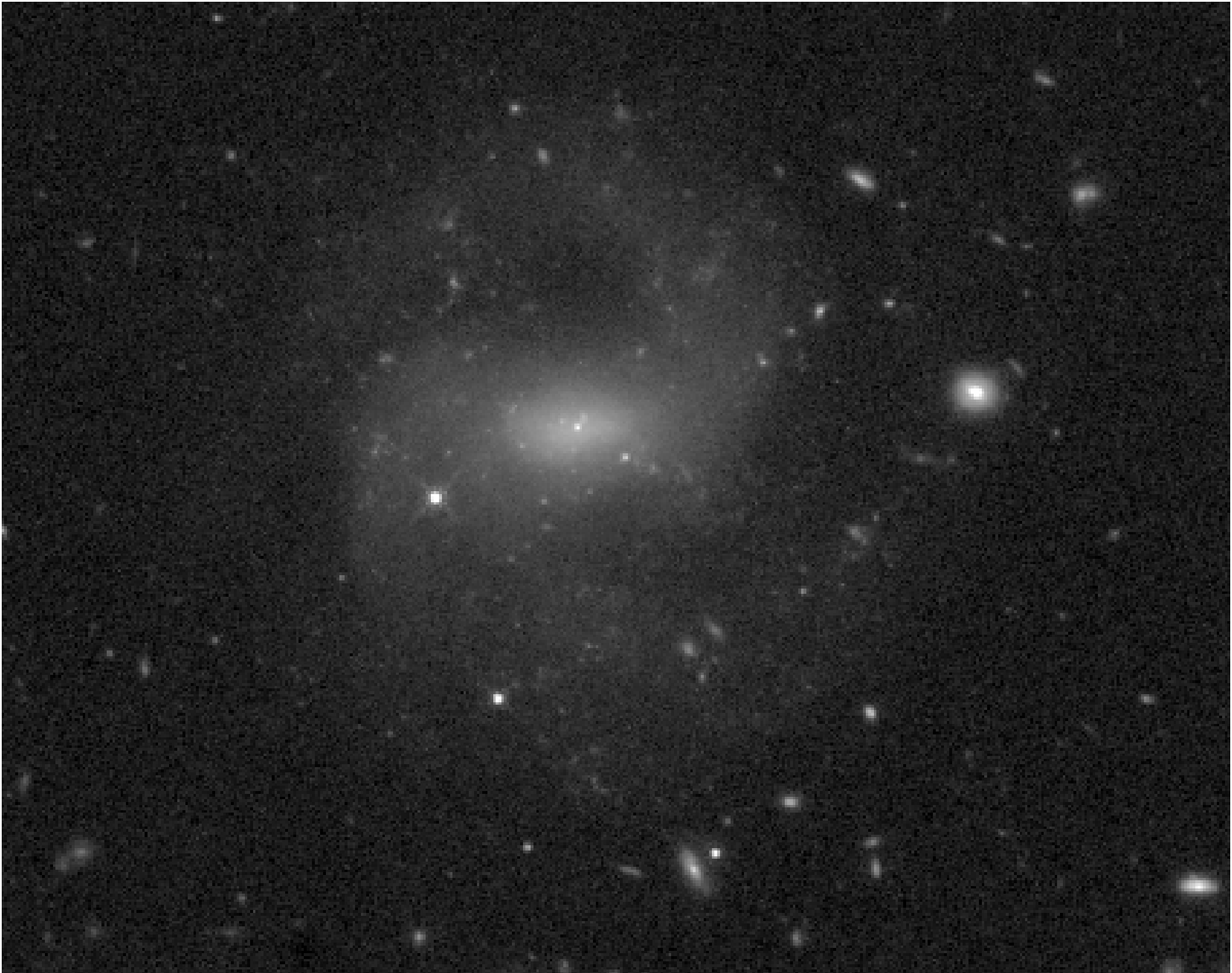}
\includegraphics[width=0.33\textwidth]{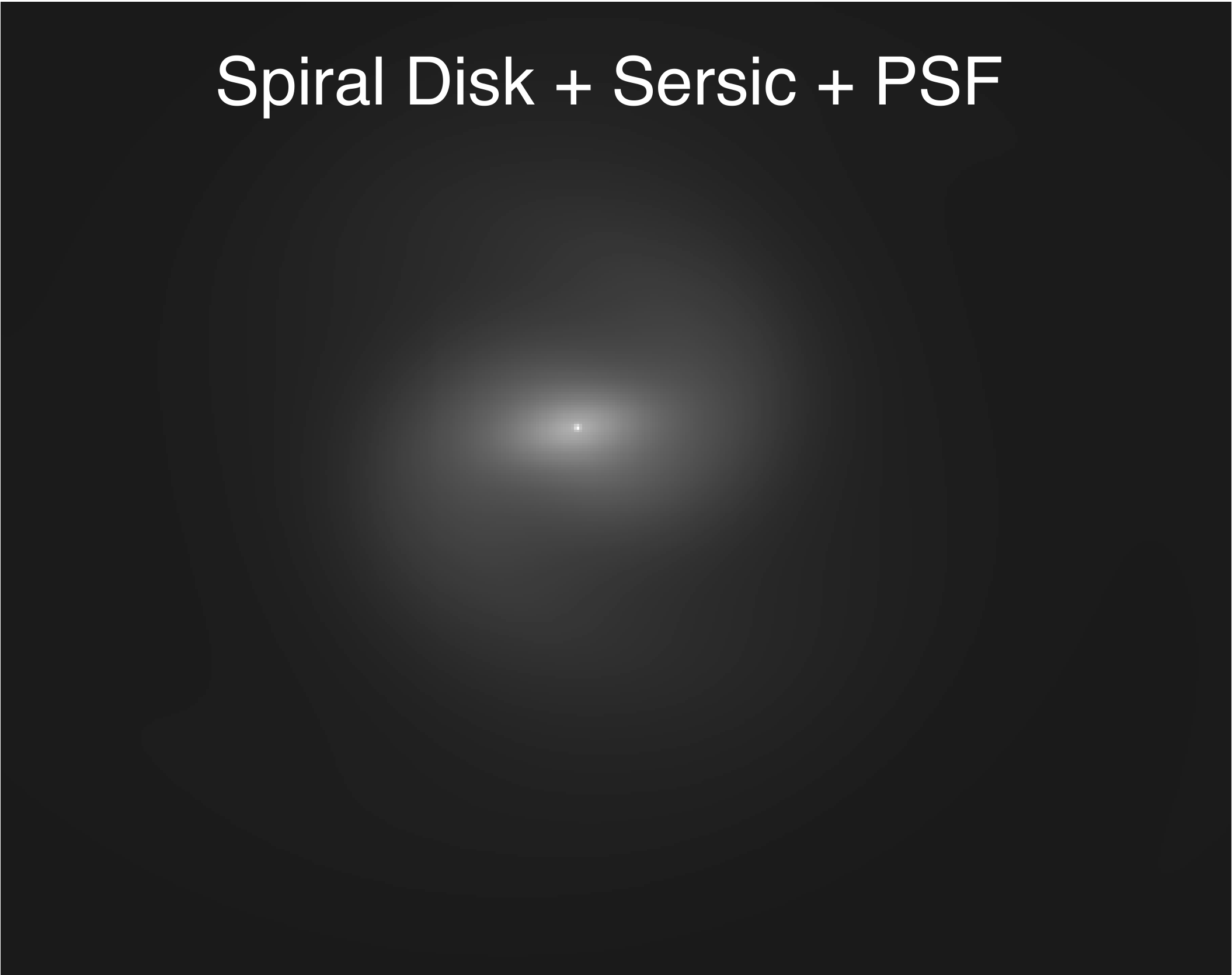}
\includegraphics[width=0.33\textwidth]{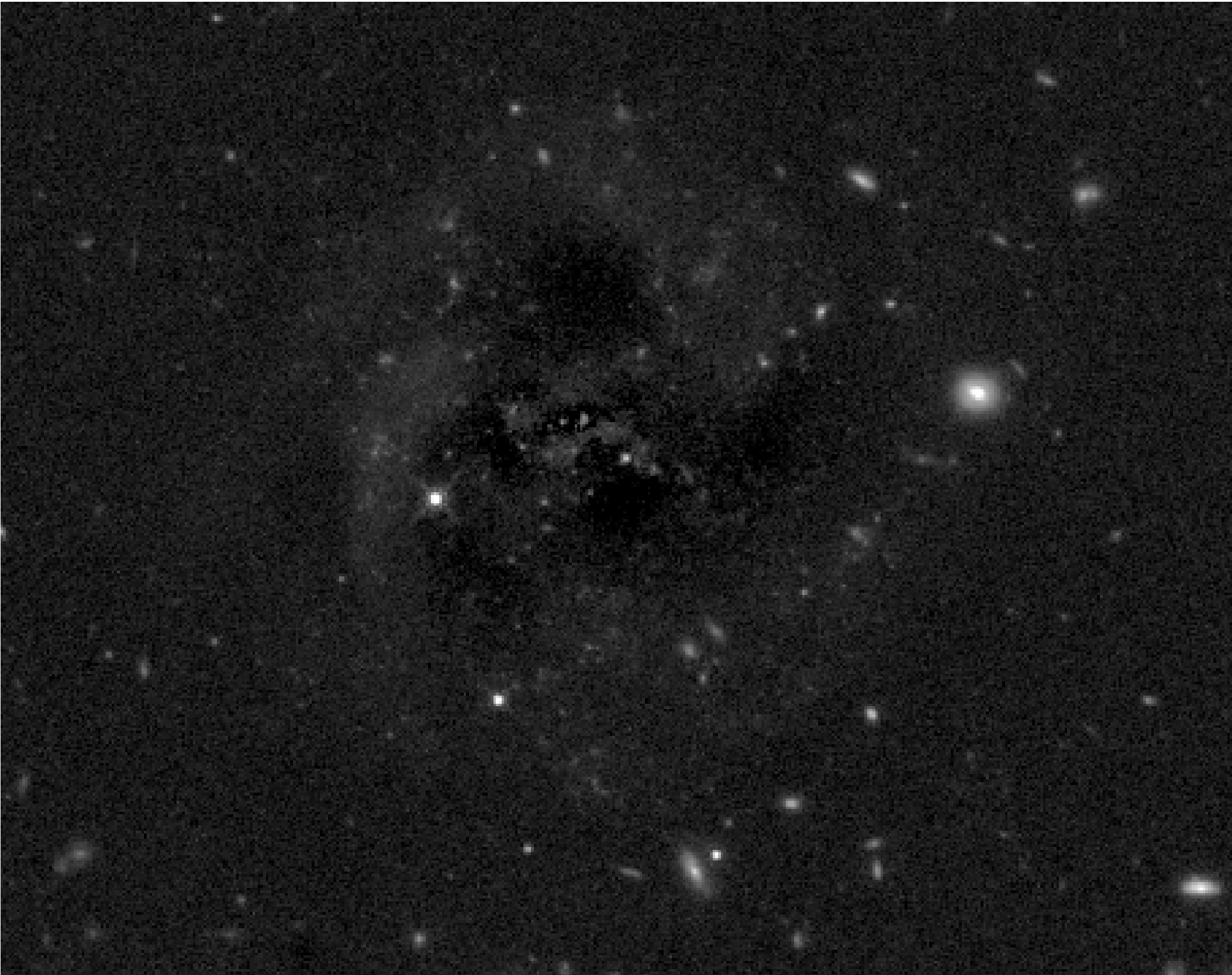}\\
\includegraphics[scale=0.6]{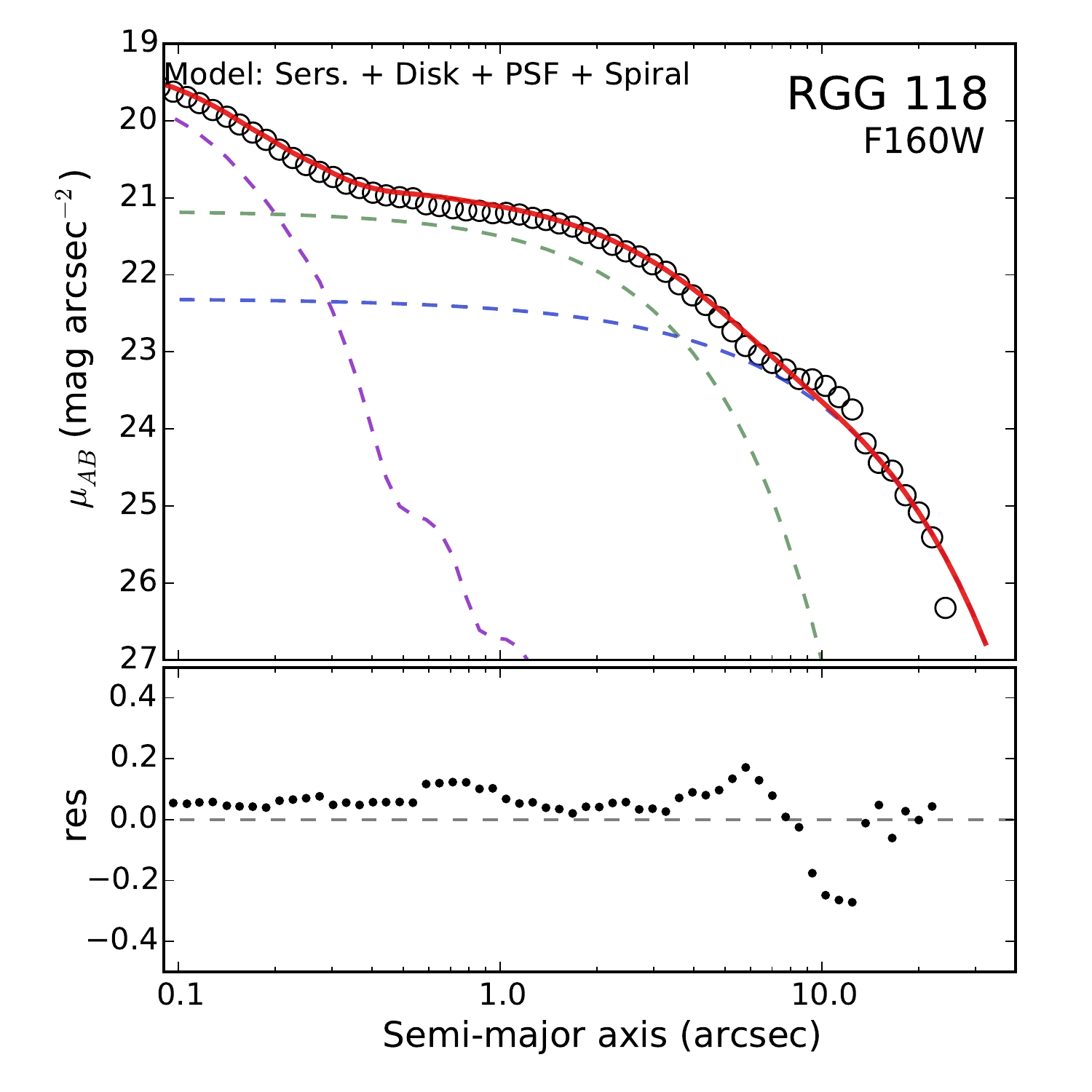}
\includegraphics[scale=0.6]{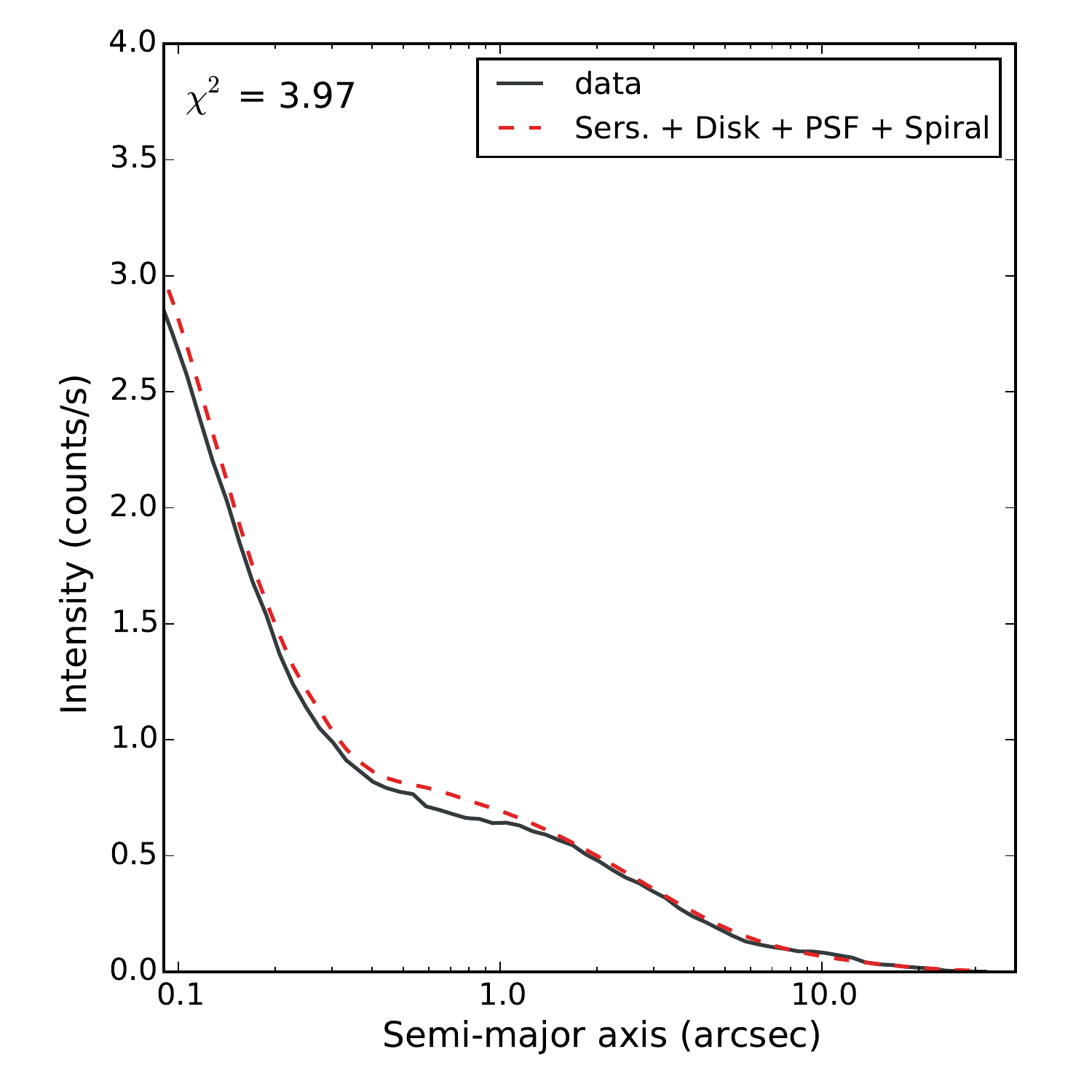}
\caption{Top row: Image of RGG 118 in the F160W filter (left); best fitting GALFIT model including a PSF, inner S{\'e}rsic component, and outer spiral disk (middle); residuals (right). Bottom row: Left panel shows the observed surface brightness profile of RGG 118 as open circles. The overall best-fit GALFIT model is shown in red, and is comprised of a PSF (purple dashed line), inner S{\'e}rsic component (green dashed line) and outer disk (blue dashed line). The residuals are shown below the surface brightness profile. Right panel shows the average intensity along a given isophote from the data and the intensity as a function of radius for the best-fit GALFIT model. Scale and colormap are consistent between the images.}
\label{SersSpPsf_}
\end{figure*}

\begin{figure*}
\includegraphics[width=0.33\textwidth]{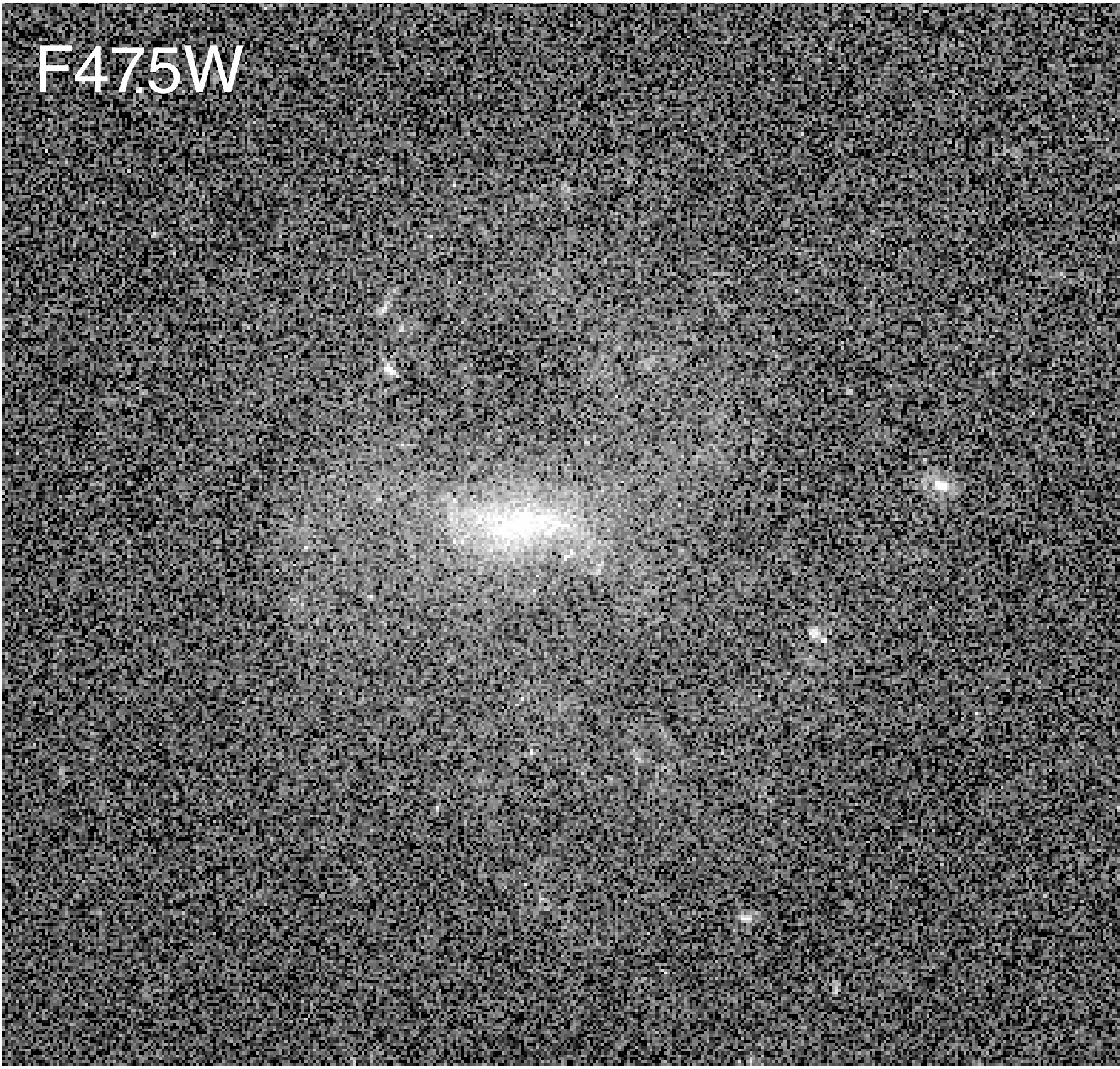}
\includegraphics[width=0.33\textwidth]{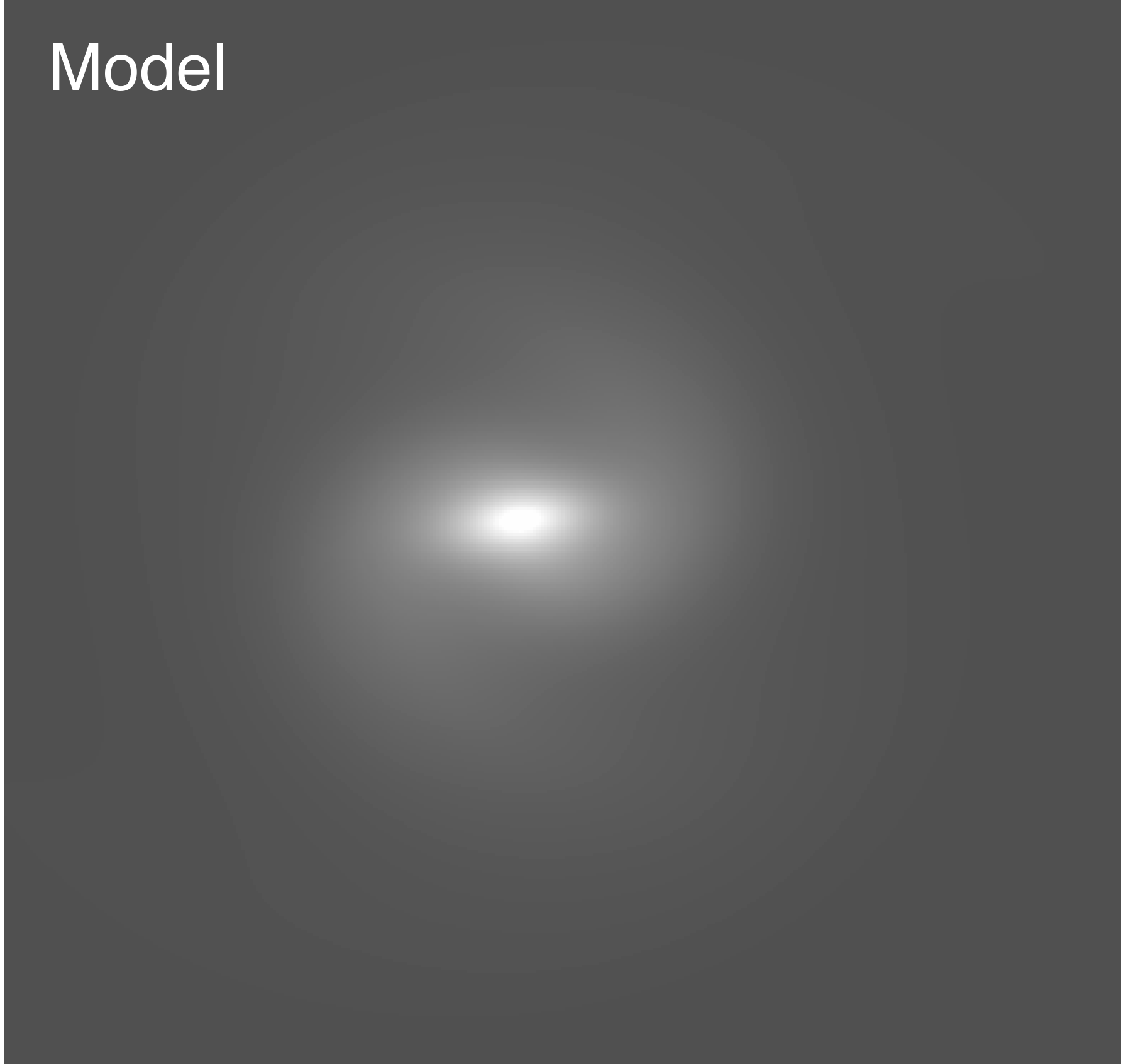}
\includegraphics[width=0.33\textwidth]{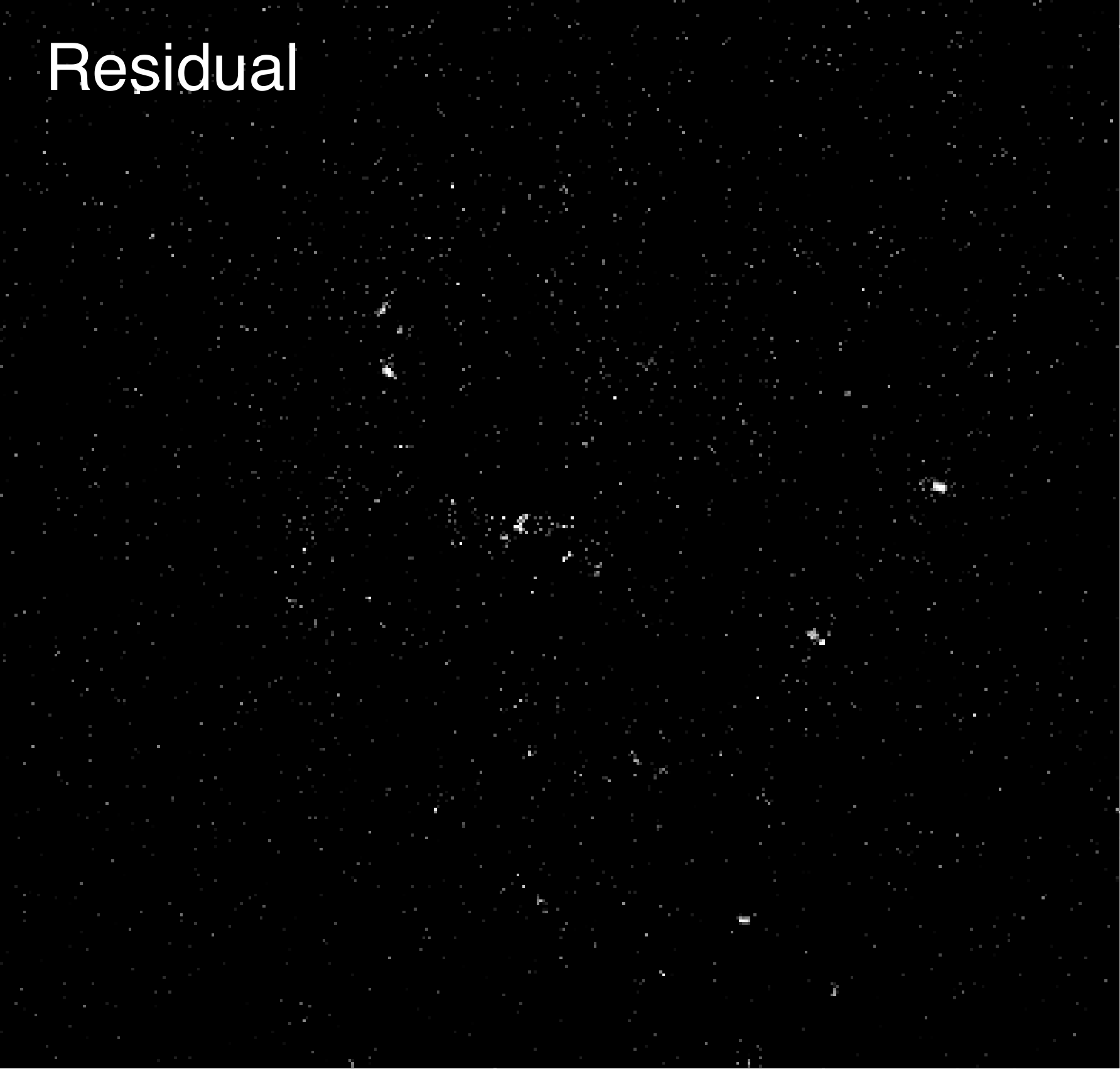}\\
\includegraphics[width=0.33\textwidth]{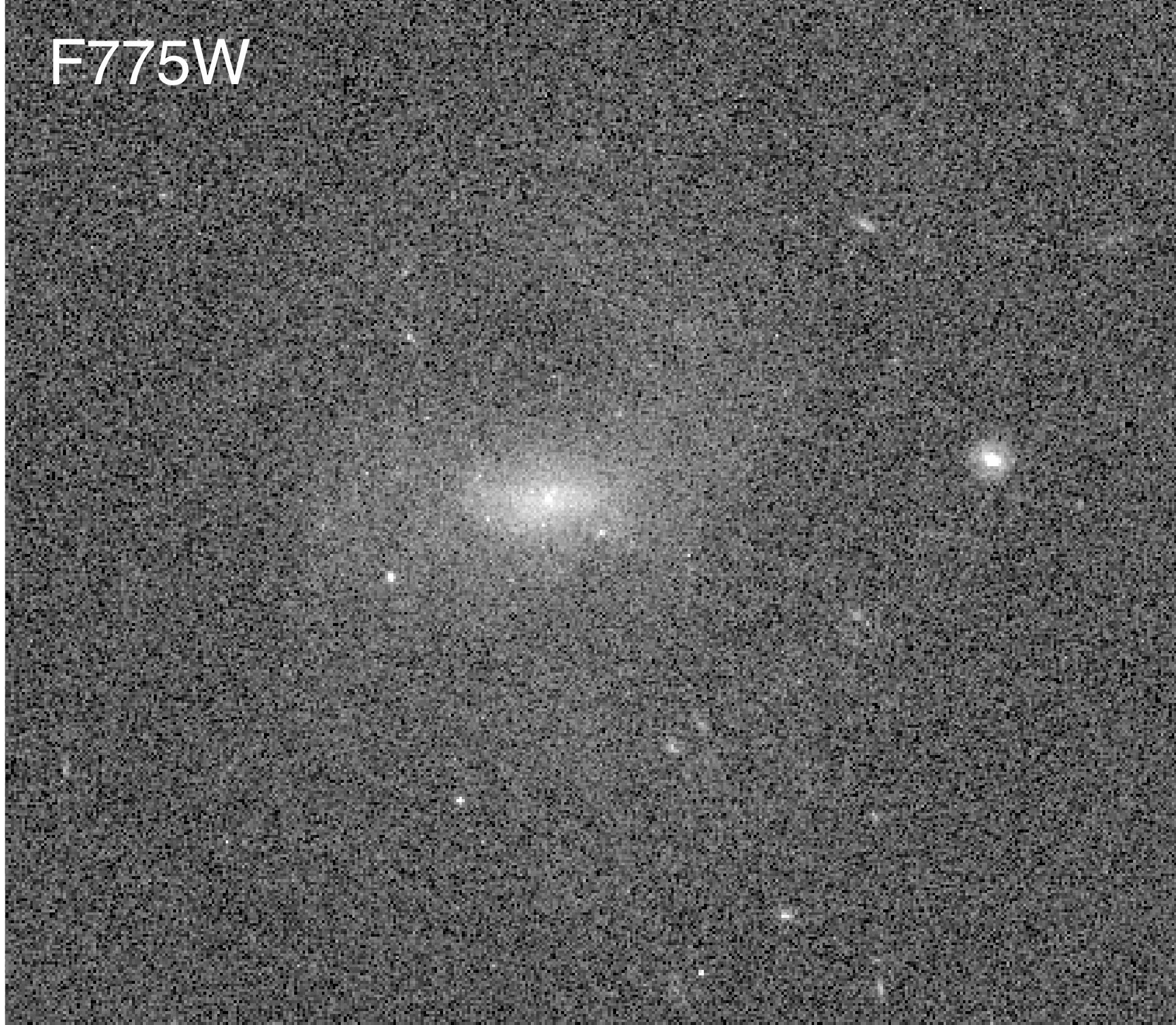}
\includegraphics[width=0.33\textwidth]{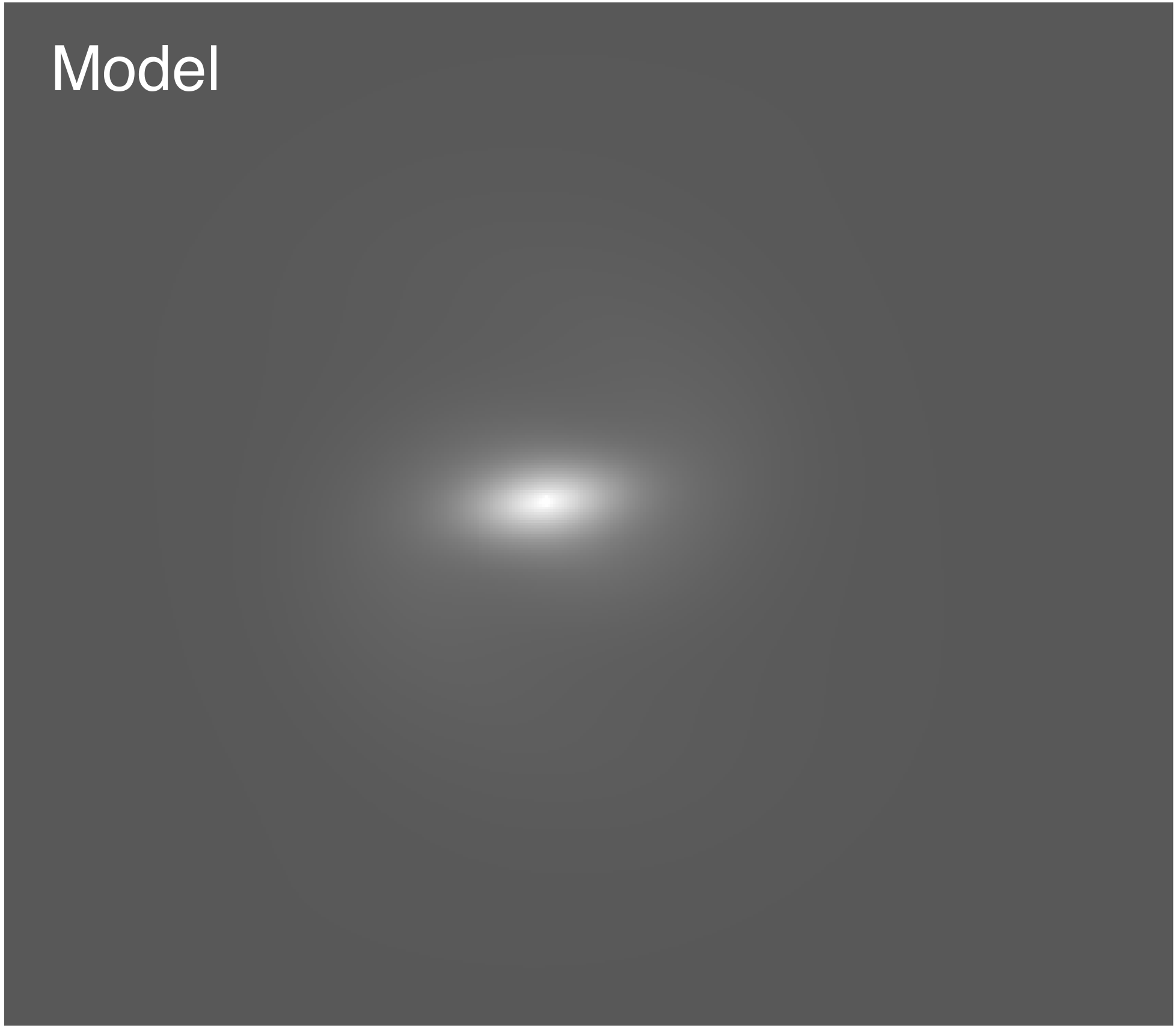}
\includegraphics[width=0.33\textwidth]{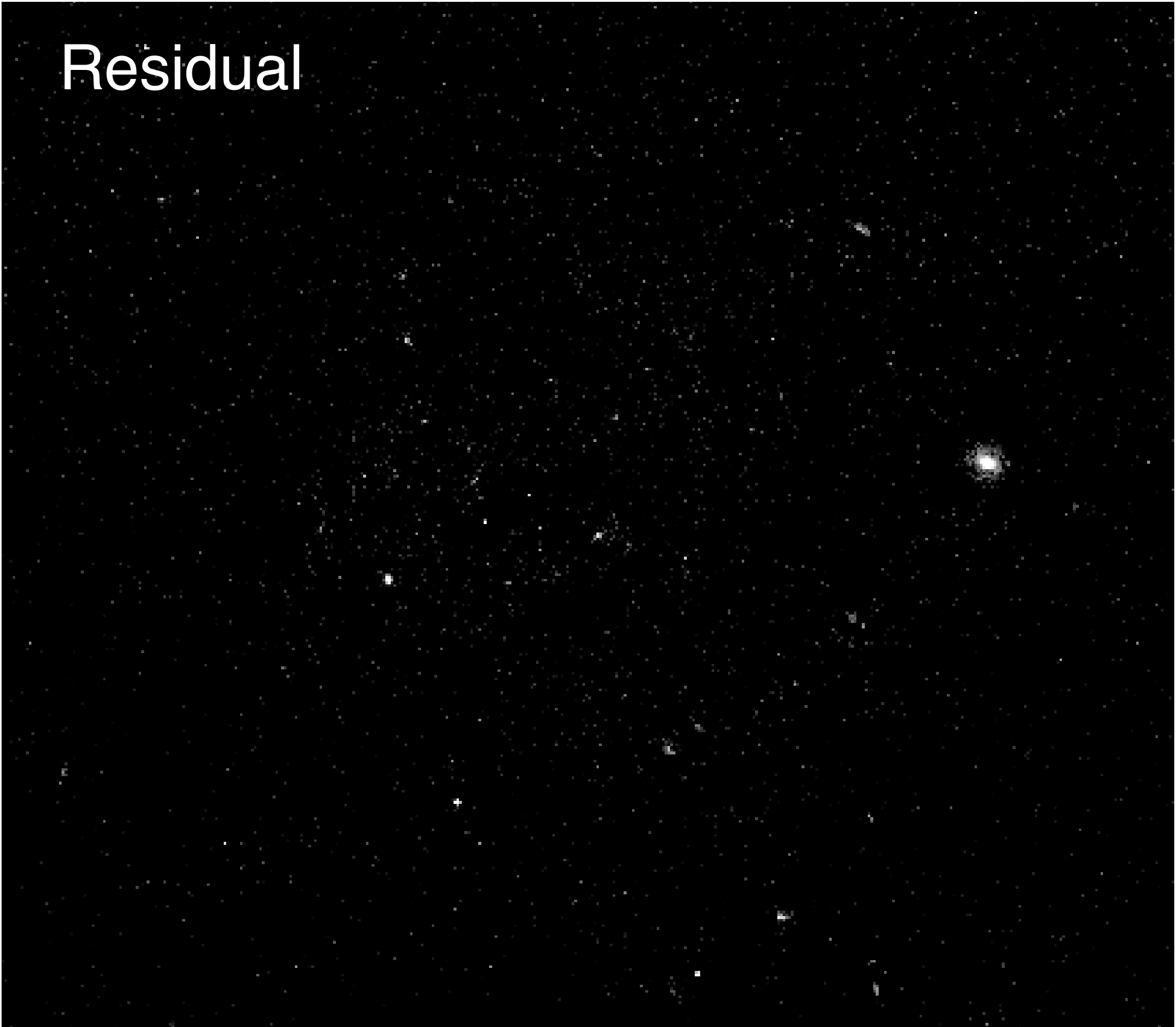}\\
\caption{Best-fit model as determined from the F160W data applied to the F475W (top row) and F775W (bottom row) images. For each filter, the scale and colormap are consistent between the images .}
\label{SersSpPsf_UVIS}
\end{figure*}

Each combination of inner and outer component is also tested with and without a central PSF component; in all cases, the inclusion of a central PSF improves the $\chi^{2}$ value by more than 40\%. Table~\ref{models} lists each tested model and its corresponding $\chi^{2}$ value, computed by comparing the intensity as a function of semi-major axis for the model and data. Ultimately, we find the best-fit model to include an outer disk ($n=1$) with spiral structure, an inner S{\'e}rsic component with $n=0.8\pm0.01$, and a central PSF (see Figure~\ref{SersSpPsf_}). The best-fit parameters (S{\'e}rsic index, effective radius, effective surface brightness) for this model are given in Table~\ref{modelparams}. Figure~\ref{SersSpPsf_UVIS} shows the S{\'e}rsic+Spiral Disk+PSF model applied to the F475W and F775W filters. In applying the model to the F475W and F775W bands, all components were held fixed except the magnitude of each component. The position of the PSF was also allowed to vary, as the angular resolution differs between the IR and UVIS filters.

The magnitudes of each component in each filter for our best-fit model are reported in Table~\ref{bestmod}. As noted in the GALFIT documentation \footnote{https://users.obs.carnegiescience.edu/peng/work/galfit/galfit.html}, the error bars returned by GALFIT rely on the assumptions that the residuals are due only to Poisson noise, and the noise has a Gaussian distribution. Similar to the procedure described in \cite{2016ApJ...823...50S}, we estimate errors on the magnitude using the standard deviation of the sky background. The standard deviation is computed by measuring the median sky value in a series of 50x50 pixel boxes placed in the sky regions surrounding the galaxy. We estimate errors on the S{\'e}rsic index and effective radii by fitting with our alternate bright star PSF (see Section 2) and taking the error to be the difference between the values of each parameter.

\floattable
\begin{deluxetable}{c c c}
\tablecaption{GALFIT fitting results \label{models}}
\tablehead{
\colhead{Components} & \colhead{$\chi^{2}$} & $M_{F160W}$ (PSF)}
\startdata
S{\'e}rsic & 347.77 & -- \\
S{\'e}rsic + PSF & 33.09 & 21.92 $\pm$0.02 \\
\hline
S{\'e}rsic + S{\'e}rsic & 353.31 & -- \\
S{\'e}rsic + S{\'e}rsic + PSF & 6.05 & 22.00$\pm$0.03 \\
S{\'e}rsic+ Disk & 281.56 & -- \\
S{\'e}rsic + Disk + PSF & 11.51 & 22.06$\pm$0.03 \\
\hline
Ferrers + S{\'e}rsic & 83.37& -- \\ 
Ferrers + S{\'e}rsic + PSF & 74.09 & 22.22 $\pm$0.03 \\
Ferrers + Disk & 82.88 & -- \\
Ferrers + Disk + PSF & 64.83 & 22.25$\pm$0.03 \\
\hline
S{\'e}rsic + Spiral Sersic + PSF & 11.46 & 22.05$\pm$0.03\\
S{\'e}rsic + Spiral Disk + PSF & 3.97 & 22.02$\pm$0.03 \\
\enddata
\tablecomments{Model components and corresponding chi-squared values for each GALFIT trial. For all models except S{\'e}rsic + Spiral Disk + PSF (shown in Figure~\ref{SersSpPsf_}. Errors on the PSF are those reported by GALFIT. }
\end{deluxetable}

\floattable
\begin{deluxetable}{c c c c | c c c c}
\tablecaption{Best fit model parameters}
\tablehead{
\multicolumn{4}{c}{Inner S{\'e}rsic component} & \multicolumn{4}{c}{Outer disk}
} 
\startdata
{${\rm r_{eff}}$} & {$\mu_{\rm eff}$} & {$n$} & (b/a) &{${\rm r_{eff}}$} & {$\mu_{\rm eff}$} & {$n$} & (b/a) \\
{(kpc)} & {(mag/arcsec$^{2}$)} & {} & & {(kpc)} & {(mag/arcsec$^{2}$)} & {}  & \\ 
\hline
$1.57\pm0.22$ & 22.5 & $0.80\pm0.1$ & 0.45 & $6.51\pm1.72$ & 24.1 & 1.00 (fixed) & 0.69 \\
\enddata
\tablecomments{Best fit model parameters (effective radius, surface brightness at the effective radius, S{\'e}rsic index, and axis ratio) for the S{\'e}rsic + Spiral Disk + PSF model in the F160W filter. }
\label{modelparams}
\end{deluxetable}

\floattable
\begin{deluxetable}{c c c c}
\tablecaption{AB magnitude of individual components \label{magnitude}}
\tablehead{
\colhead{Filter} & \colhead{PSF} & \colhead{Inner component} & \colhead{Disk}\\
\colhead{} & \colhead{(mag)} & \colhead{(mag)} & \colhead{(mag)}
} 
\startdata
F475W & 22.56$^{+0.37}_{-0.28}$ & 19.38$^{+0.36}_{-0.27}$ & 17.49$^{+0.65}_{-0.40}$\\
F775W  & 22.37$^{+0.38}_{-0.28}$  & 18.20$^{+0.13}_{-0.12}$ & 17.48$^{+0.82}_{-0.46}$ \\
F160W & 22.02$\pm$0.03 & 18.31$^{+0.09}_{-0.08}$ & 16.42$^{+0.14}_{-0.12}$ \\
\enddata
\tablecomments{AB magnitude of each component in the best-fitted model (S{\'e}rsic + Spiral Disk + PSF) for each filter. Modeling was performed on the F160W filter, and the best-fitted model was applied to the two optical filters. }
\label{bestmod}
\end{deluxetable}

\subsection{Colors and stellar masses}

Using the $g$, $i$, and $H$-band magnitudes from GALFIT and extinction corrections based on the extinction map from \cite{2011ApJ...737..103S}, we find the $g-i$ color and $H$-band luminosity for the outer disk and inner S{\'e}rsic components. For the inner component, we find $(g-i)_{\rm bulge} = 1.18^{+0.48}_{-0.40}$. 
The disk is faint in the g and i bands, and the errors on the disk magnitudes returned from GALFIT are large (they give $(g-i)_{\rm disk} =-0.01^{+1.1}_{-1.2}$). An alternate way to try to constrain the disk color is using the 1-D light profiles output by \textit{ellipse}. Using the total flux computed between a radius of 10$''$ and 16$''$, i.e., where the disk is dominant, we find $(g-i)_{\rm disk}=0.5\pm0.3$.

\begin{figure}
\includegraphics[width=0.45\textwidth]{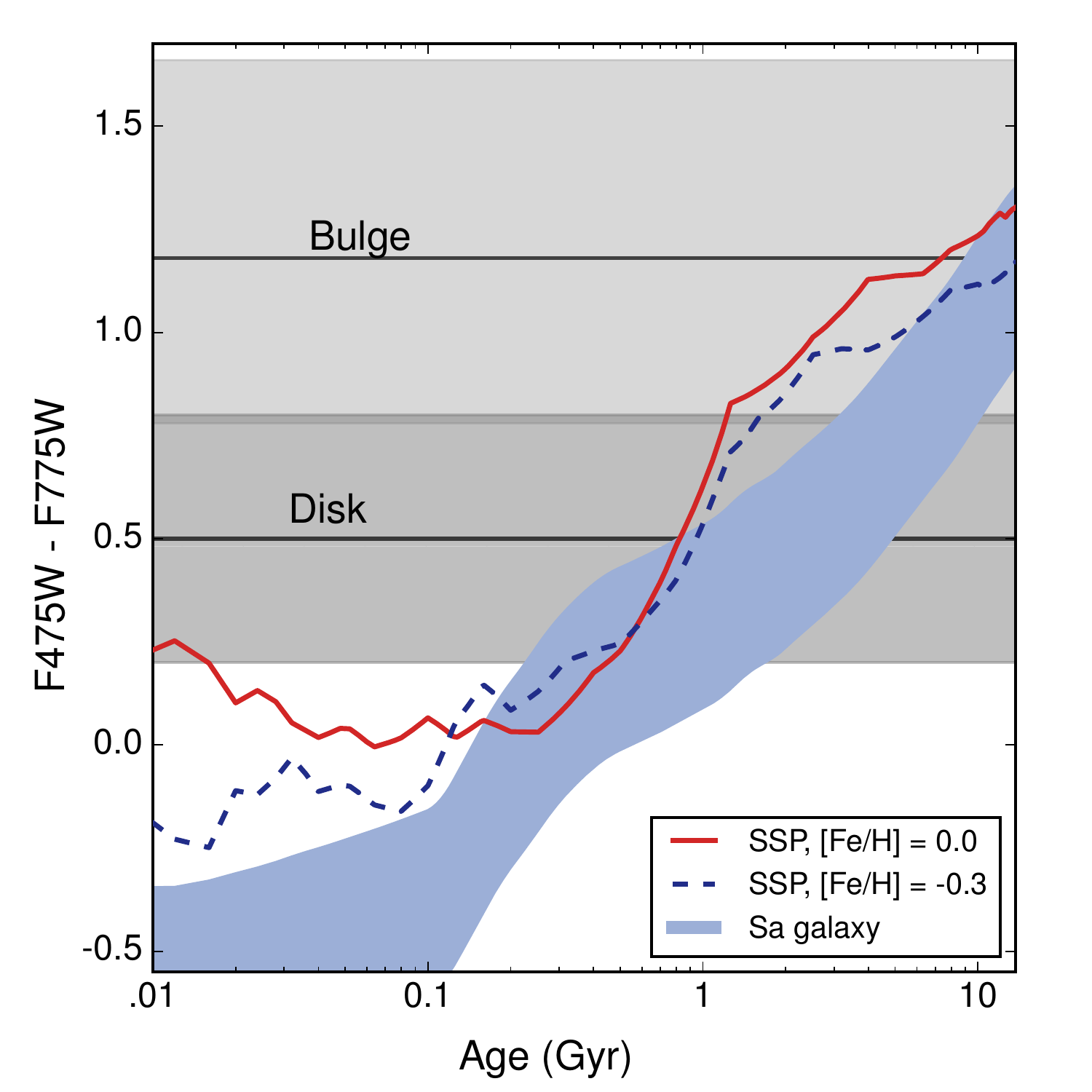}
\caption{F475W - F775W color versus stellar population age. The solid red and dashed blue lines represent models for the extinction-corrected color evolution of a single stellar population with initial stellar mass of $10^{8}M_{\odot}$. The solid red line shows the evolution of a population with solar metallicity, while the dashed blue shows a population with sub-solar metallicity ($[{\rm Fe/H}] =-0.3$, or roughly half the metallicity of the Sun). The gray horizontal lines show the color of the bulge and disk, and the corresponding shaded regions encompass the errors. Note that the disk color is computed between a radius of 10$''$ and 16$''$. The light blue shaded region encompasses evolutionary tracks for an Sa-galaxy (the morphological classification most similar to RGG 118) with a \textit{total} mass of $5\times10^{9}M_{\odot}$ and chemically consistent metallicity. Models for this galaxy were computed for E(B-V) ranging from 0.0 to 0.5; increasing E(B-V) increases (reddens) the F475W-F775W color. All evolutionary tracks were computed using GALEV \citep{Kotulla:2009ul}. }
\label{ssp}
\end{figure}

We show the $g-i$ (F475W-F775W) color evolution of single stellar population with initial mass of $10^{8}$ solar masses using GALEV \citep{Kotulla:2009ul} and show the results in Figure~\ref{ssp} for a solar-metallicity model and sub-solar metallicity model. While treating the bulge and disk as single stellar populations is a significant simplification, we can nevertheless get a rough idea of the relative ages of the bulge and disk. A disk with $(g-i)\approx0.5$ would be dominated by young stellar populations with ages of hundreds of Myr to $\sim1$ Gyr. The observed bulge color suggests a population that is older than $\sim1$ Gyr. We also show the color evolution for an Sa-galaxy with total mass of $5\times10^{9}M_{\odot}$ in Figure~\ref{ssp}.

We compute the stellar mass of each component using the color-based mass-to-light ratios derived by \cite{2003ApJS..149..289B}. We measure the luminosity using our F160W observations (roughly the equivalent of $H$ band), since the variation in M/L is decreased at NIR wavelengths. We use the $(g-i)$ color to compute the $\log({\rm M/L}_{H})$ ratio $\Upsilon_{H}$. The resulting equation is $\Upsilon_{H} = -0.186 + (0.179\times(g-i))$. We compute a disk stellar mass of $M_{\ast, disk} = 10^{9.23(+0.1,-0.09)}M_{\odot}$ and an inner component stellar mass of $M_{\ast, bulge} = 10^{8.59(+0.11,-0.12)}M_{\odot}$. This gives a total stellar mass of $M_{\ast, total} = 10^{9.32(+0.10,-0.11)}M_{\odot}$. This is in good agreement with the total stellar mass in the NASA-Sloan Atlas, which uses the k-correct code \citep{2007AJ....133..734B}: $M_{\ast, total} = 10^{9.35} M_{\odot}$.

\section{Discussion}

\subsection{Nature of the central point source}

In the following section, we discuss the nature of the observed point source. We first consider whether the optical point source is consistent with an AGN given the X-ray luminosity and \textit{assuming a typical quasar SED}. Using the quasar SED from \cite{2006ApJS..166..470R}, we use the observed X-ray luminosity (from \citealt{2015ApJ...809L..14B})  to determine the expected luminosity at the central wavelength in the F475W filter. The \cite{2006ApJS..166..470R} SED is computed out to a maximum energy 0.4 keV, beyond which it is assumed to have constant $\nu L_{\nu}$. From our X-ray observations, $\rm{\nu L_{\nu}} (2 \rm{keV})$ = $1.96\times10^{39}$ \ergs. Based on the \cite{2006ApJS..166..470R} SED, we expect $\rm{\nu L_{\nu}} (4659 \rm{\AA})$ = $2.7\times10^{40}$ \ergs. 
The measured luminosity of the point source is $\rm{\nu L_{\nu}} (4659 \rm{\AA})$ = $2.5\times10^{40}$ \ergs, based on the magnitude of the point source as determined in GALFIT (22.56 mag). While we note that there is considerable scatter ($\sim0.5$ dex) in the \cite{2006ApJS..166..470R} mean quasar SED, the measured luminosity is in excellent agreement with the predicted luminosity, suggesting the point source is indeed dominated by the AGN. 

We also consider the possibility of a nuclear star cluster (NSC) for the point source. NSCs become increasingly prevalent as one moves down the galaxy mass function, with as many as $80\%$ of galaxies with $M_{\ast}<10^{10}M_{\odot}$ hosting a massive, compact NSC (see, e.g., \citealt{Carollo:1997fj, Carollo:1998uq, 2002AJ....123.1389B, 2006ApJS..165...57C, 2007ApJ...671.1456C}). These NSCs are typically a few to a few tens of parsecs in radius, with masses from $\sim10^{5}-10^{7}M_{\odot}$ \citep{2002AJ....123.1389B, 2004AJ....127..105B, 2005ApJ...618..237W}. At the distance of RGG 118, 0.1$''$ corresponds to 50 pc, meaning that for our observations, a NSC would appear as an unresolved point source.

\textit{HST} surveys of nearby late-type galaxies have revealed relations between galaxy properties and those of their NSCs (see \citealt{2002AJ....123.1389B, 2004AJ....127..105B}). In particular, \cite{2004AJ....127..105B} finds a relation between the B-band magnitude of the galaxy and the I-band magnitude of the NSC. RGG 118 has an absolute B-band magnitude of $M_{B, \rm{galaxy}} = -18.39\pm0.521$ (from the HyperLeda database; \citealt{2014AA...570A..13M}). Using the relationship from Table 2 in \cite{2004AJ....127..105B}, we find that, if RGG 118 has a NSC, it's predicted I-band magnitude is $M_{I, \rm{NSC}} = -11.74$. Using the synthetic photometry package SYNPHOT, we compute the predicted WFC3 F475W and F775W magnitudes. We assume a stellar population with an age of 1 Gyr and star formation occurring in an instantaneous burst, normalized to the predicted I-band magnitude. With these assumptions, SYNPHOT predicts NSC apparent magnitudes of $m_{F475W, \rm{NSC}} = 24.4$ and  $m_{F775W,\rm{NSC}} = 23.7$; one to two magnitudes fainter than the observed point source in RGG 118. 

Given the mass and morphology of RGG 118, it is possible that it does contain a NSC. Based on scaling relations between galaxy stellar mass and NSC properties \citep{2016MNRAS.457.2122G}, the NSC would have an expected radius of $\sim2-3$ pc, and the combined BH+NSC mass would be $\sim8\times10^{5}M_{\odot}$. However, while it may contain a NSC, the point source luminosity in RGG 118 is consistent with being dominated by the AGN. We note that our main results relating to the structure of RGG 118 are unaffected by the relative contributions to the central PSF from an AGN versus a NSC.

\subsection{Comparison to SDSS imaging analysis}

\cite{2015ApJ...809L..14B} first analyzed the SDSS imaging of RGG 118. They also used GALFIT to decompose the SDSS image into individual components, finding a best-fitted model including an exponential disk, inner S{\'e}rsic component with $n=1.13\pm0.26$, and PSF. The masses of the disk and inner component were found to be $10^{9.3}\pm0.1M_{\odot}$ and $10^{8.8\pm0.2}M_{\odot}$, consistent with the masses determined in this work. We do find a slightly lower S{\'e}rsic index for the inner component based on modeling of the HST data than for the SDSS data ($n=0.8\pm0.1$ in this work compared to $n=1.13\pm0.26$). 

The SDSS imaging was subsequently analyzed by \cite{2016ApJ...818..172G}. They use 1-D modeling techniques to model the profile of RGG 118.  \cite{2016ApJ...818..172G} include spheroid (bulge), bar, and disk components, and find that a central PSF is not required for their model. Their bar component is fit with a modified Ferrers profile, while their bulge is fit with a S{\'e}rsic component of n=0.41. Our attempts to model the RGG 118 light profile with a bulge, bar, and disk did not converge on a solution in GALFIT.  

One potential explanation for our differing preferred models is that their spheroid component has an effective radius of 0.63$''$, less than the typical FWHM of the SDSS r-band PSF (1.3$''$), making it difficult to distinguish their bulge from a point source. We are able to find best-fit solutions for models including a point source, bar, and disk, though these have higher $\chi^{2}$ values than models without a bar (Table~\ref{models}). \cite{2016ApJ...818..172G} find the stellar masses of the disk, bar, and bulge to be $10^{9.36}M_{\odot}$, $10^{7.76}M_{\odot}$, and $10^{7.92}M_{\odot}$, respectively. Their total stellar mass is consistent with our findings, though the masses of the individual components are not. Overall, we find models including Ferrers (bar) components to produce poorer fits to the HST data.

\begin{figure}
\includegraphics[width=0.5\textwidth]{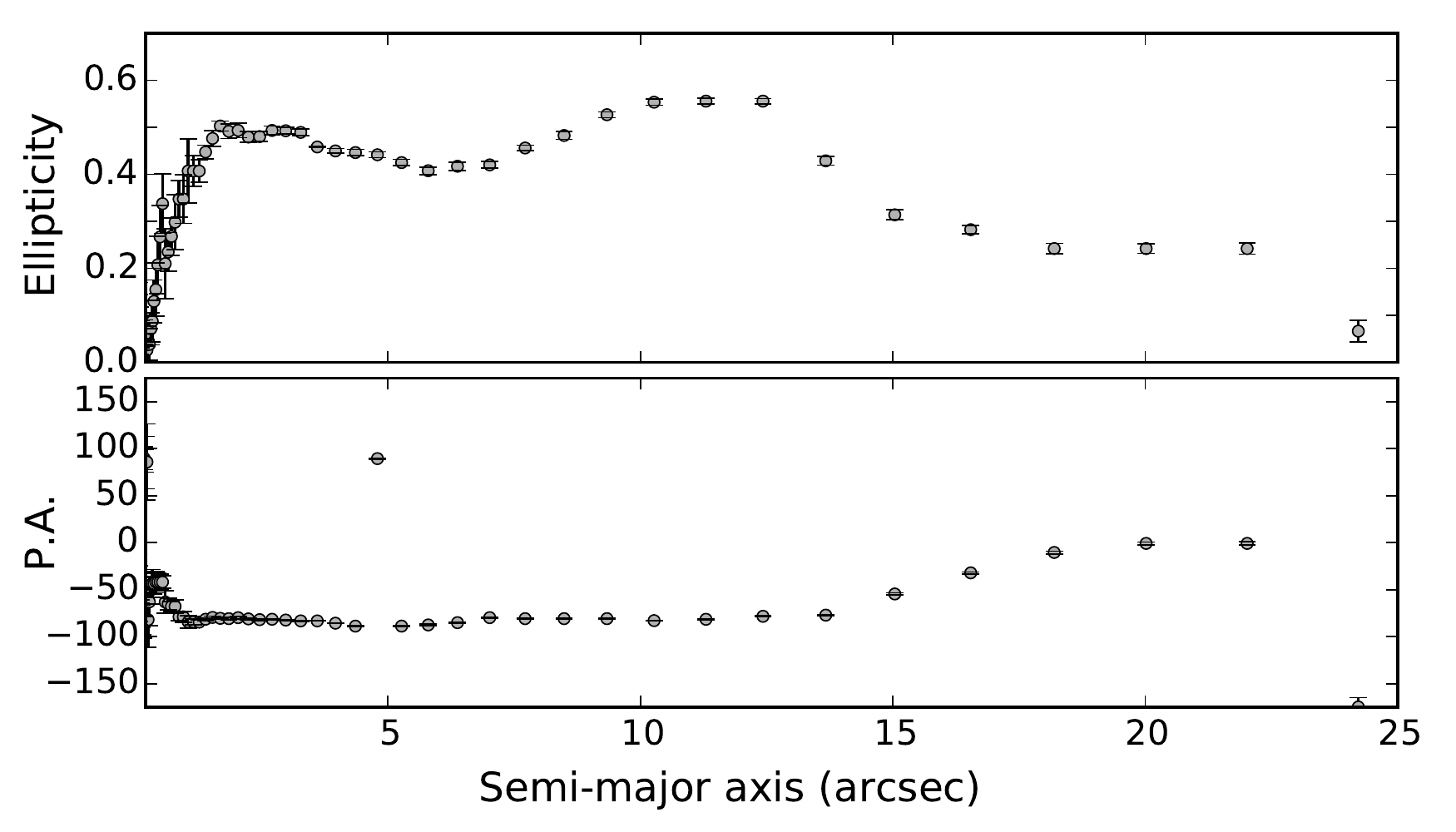}
\caption{Position angle and ellipticity profile of RGG 118. Signatures of a bar are an increase in ellipticity with fixed position angle within the bar, followed by a  sharp drop off in ellipticity coincident with a change in position angle. We do not observe these features for RGG 118.}
\label{pa_ellip}
\end{figure}

The presence of a bar can also be assessed using the ellipticity and position angle profiles of RGG 118 (Figure~\ref{pa_ellip}). \cite{2007ApJ...657..790M} describe several signatures produced by a bar. Within the bar, there is typically a continuous increase in ellipticity with a fixed position angle. At the end of the bar, there is an abrupt drop-off in ellipticity and a sharp change in position angle as the profile moves from bar-dominated to being dominated by the disk. We do not find evidence for a bar in either the ellipticity or position angle profile of RGG 118.

\subsection{Comparison to other systems}

Studying the morphologies of the population of dwarf/low-mass galaxies with AGNs may help illuminate what factors are important for influencing the presence of an AGN in these systems. Here, we compare RGG 118 to other low-mass galaxies with AGN, as well as to the general population of spiral galaxies. \cite{2011ApJ...742...68J} study the structures of 147 host galaxies of low-mass AGNs ($M_{\rm BH}\lesssim10^{6}~M_{\odot}$), the vast majority of which have extended disks. They find that for galaxies with detected disks, the mean ratio of the bulge-to-total luminosity $\langle B/T \rangle$ is 0.23 (with a median of 0.16). We find that the $B/T$ ratio for RGG 118 is 0.15$\pm0.03$. 

We can also compare RGG 118 to disk-dominated galaxies studied by \cite{2003ApJ...582..689M}, who presented bulge-to-disk decompositions of 121 late-type spiral galaxies. They found that the bulge S{\'e}rsic indices ranged from 0.2 - 2.0, with a mean of $\sim1.0$. They also found a relation between the bulge and disk radii, such that the mean ratio $\langle r_{e}/r_{h} \rangle = 0.22\pm0.09$. RGG 118 is consistent with these disk-dominated galaxies, with a bulge S{\'e}rsic index of $n=0.8$ and a bulge-to-disk scale length ratio of 0.24. 

There are also examples of low-mass galaxies with AGNs that have very different morphologies from RGG 118. For example, NGC 4395 has a disk and nuclear star cluster, but is bulgeless \cite{2003ApJ...588L..13F}. POX 52 on the other hand has no detected disk component and has a S{\'e}rsic index of $n=4.0$ \citep{2008ApJ...686..892T}. Accreting BHs have been found in the compact irregular dwarf galaxy Henize 2-10 \citep{Reines:2011fr, 2012ApJ...750L..24R, 2016ApJ...830L..35R}, and in a member of the the interacting dwarf galaxy pair Mrk 709 \citep{2014ApJ...787L..30R}. Further demographic studies will be necessary to determine whether the morphology of dwarf galaxies with AGN is distinct from those without.

\subsection{Scaling relations}

There are well known scaling relations between BH mass and bulge properties such as stellar velocity dispersion \citep{2000ApJ...539L...9F, 2000ApJ...539L..13G, 2009ApJ...698..198G, 2013ApJ...764..184M}, stellar mass \citep{Marconi:2003fk, Haring:2004lr}, and near-infrared luminosity \citep{Marconi:2003fk}. These relations imply that the BH and galaxy co-evolve despite the small gravitational sphere of influence of the BH relative to the galaxy. In this section, we discuss the importance of constraining the low-mass end of scaling relations and revisit the position of RGG 118 relative to these relations. 

Cosmological simulations suggest that the BH occupation fraction in low-mass galaxies, as well as the slope and scatter of the low-mass end of BH-galaxy scaling relations are related to the primary mechanism by which BH seeds form in the early universe \citep{2009MNRAS.400.1911V}. BH seed formation models tend to fall into two categories: light seeds ($M_{\rm BH, seed}\approx100~M_{\odot}$) and heavy seeds ($M_{\rm BH, seed}\approx10^{4-5}~M_{\odot}$). In light seed models, BH seeds form from the deaths of Population III stars \citep{2001ApJ...551L..27M,2009ApJ...701L.133A, 2014ApJ...784L..38M}. These models predict a plume of objects which scatter \textit{below} the present-day M-$\sigma$ relation at low galaxy/BH masses. On the other hand, heavy seed models \citep{2006MNRAS.370..289B, 2006MNRAS.371.1813L} produce BH seeds via direct collapse of gas clouds and predict that objects at the low-mass end of M-$\sigma$ should scatter \textit{above} the relation. 

There has also been considerable discussion regarding whether galaxies without classical bulges follow scaling relations. \cite{Kormendy:2013ve} find that the properties of galaxies with pseudobulges (i.e., flatter, rotationally supported components with S{\'e}rsic indices $n<2.0$; \citealt{2004ARAA..42..603K}) do not correlate with BH mass. Though there is considerable intrinsic scatter in these scaling relations, it seems that galaxies with pseudobulges tend to fall below the $M_{\rm BH}-M_{\rm bulge}$ relation, such that their BHs are under-massive with respect to the mass of the (pseudo)bulge \citep{Greene:2010fr}. With a S{\'e}rsic index of $n=0.8$, the central component of RGG 118 is more consistent with a pseudobulge than a classical bulge. \cite{2015ApJ...809L..14B} find that RGG 118 does sit below the $M_{\rm BH}-M_{\rm bulge}$ relation defined by early-type galaxies; our results based on \textit{HST} imaging are consistent with this. For the bulge mass of RGG~118 ($10^{8.59}M_{\odot}$), the relation given by \cite{Kormendy:2013ve} predicts a BH mass of $\sim8\times10^{5}M_{\odot}$, or 0.2\% of the bulge mass. The BH in RGG 118 is in actuality roughly an order of magnitude smaller. We also find that RGG 118 sits below the relation between BH mass and IR bulge luminosity. 

\begin{figure*}
\centering
\includegraphics[width=0.44\textwidth]{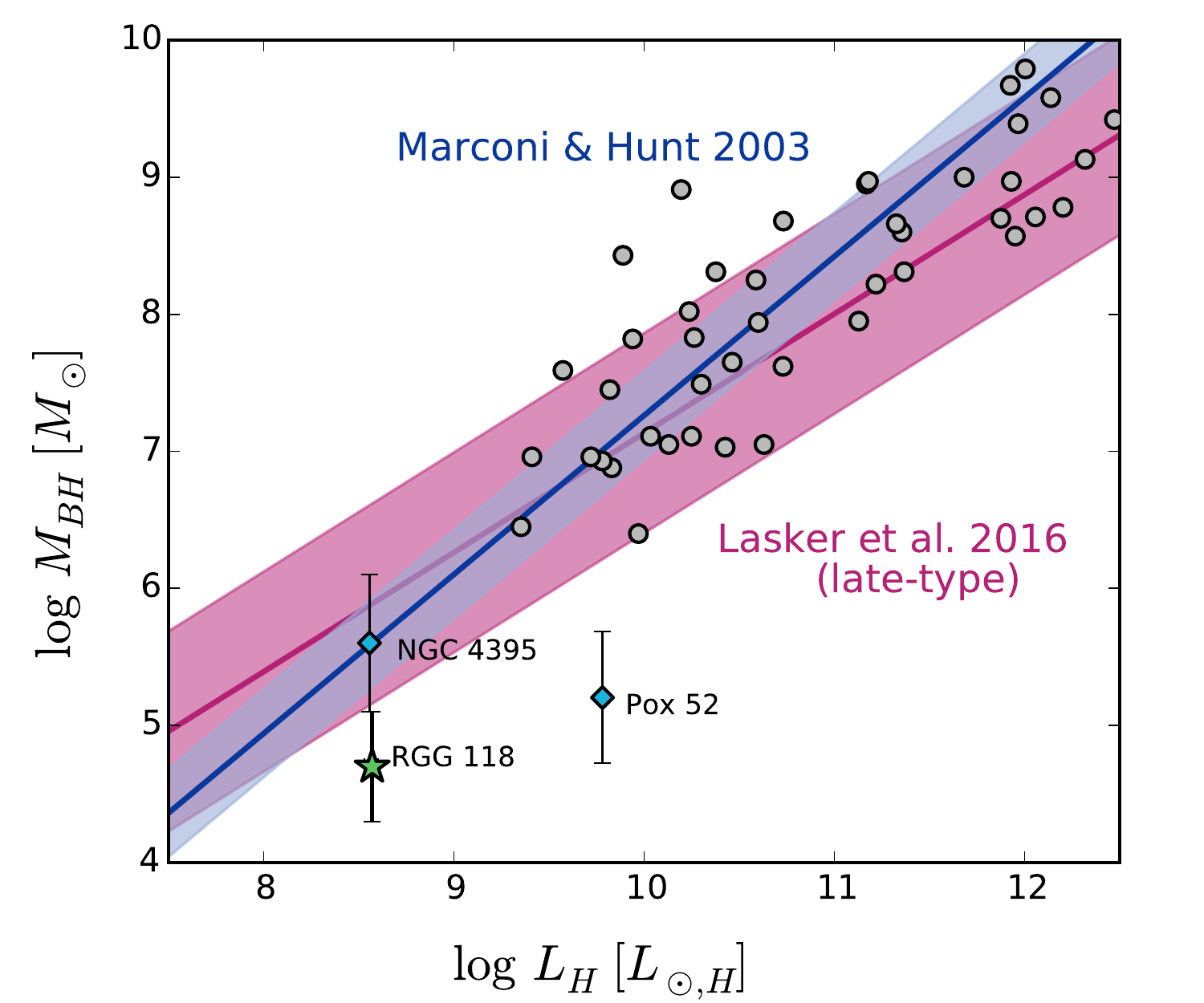} \includegraphics[width=0.44\textwidth]{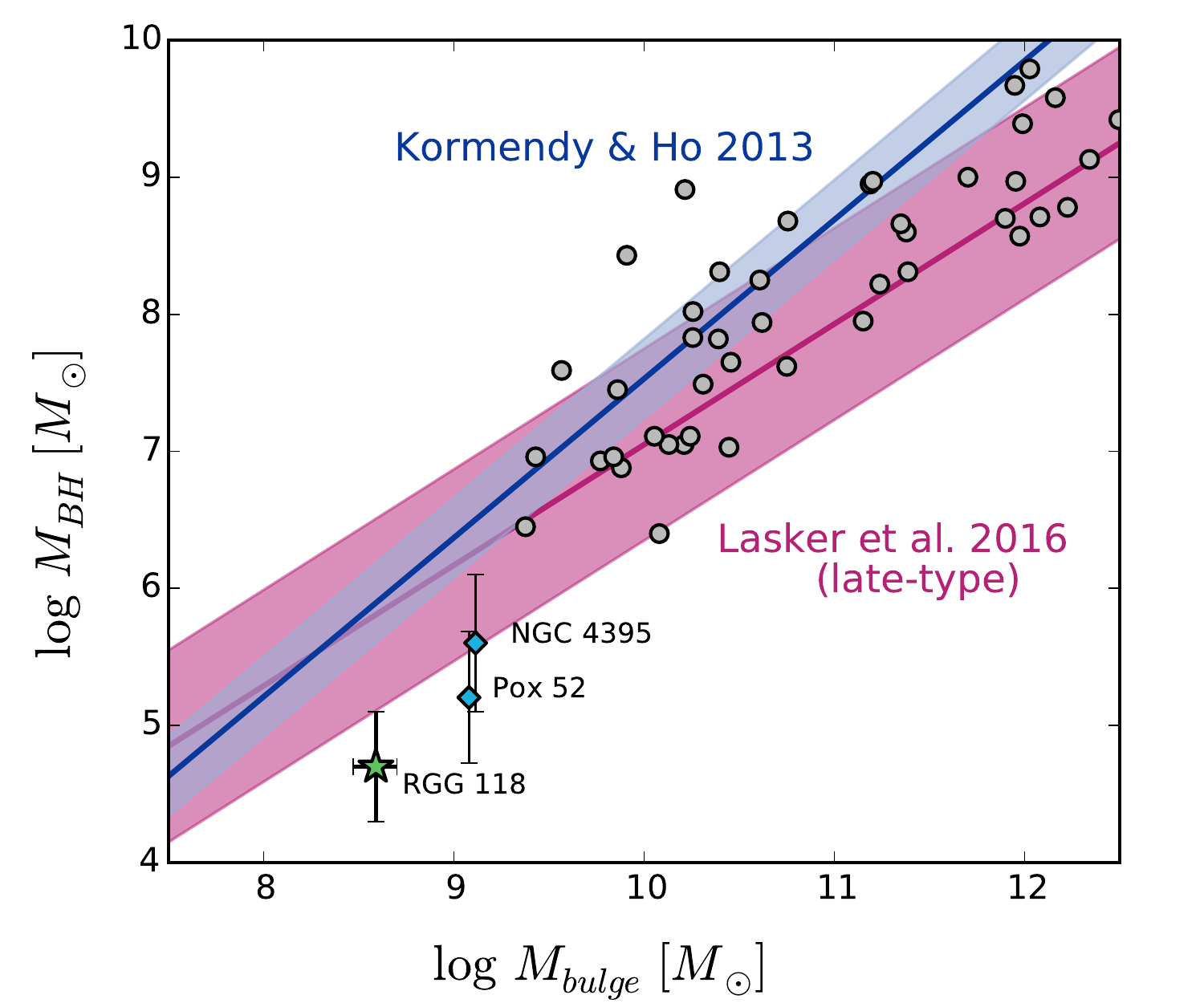}
\caption{Scaling relations between bulge properties and BH mass. \textit{Both panels:} The green star represents RGG 118, and the gray circles show galaxies from \cite{2014ApJ...780...70L} and \cite{2016ApJ...825....3L}. We also show the positions of NGC 4395 \citep{2003ApJ...588L..13F, 2015ApJ...809..101D} and Pox 52 \citep{2004ApJ...607...90B, 2008ApJ...686..892T} . Note that NGC 4395 is bulgeless, so the ``bulge" stellar mass and luminosity refer to the entire galaxy. WFC3 H-band luminosities for NGC 4395 and Pox 52 are computed by transforming their 2MASS H-band luminosities via the relations given in \cite{2011wfc..rept...15R}. The pink lines and shading show the scaling relations derived by \cite{2016ApJ...825....3L} including the offsets found for their late-type galaxy sample relative to the full sample. \textit{Left:} $M_{BH}$ versus $L_{\rm bulge, H}$. The blue line and shaded region show the $L_{\rm bulge}-M_{BH}$ relation and intrinsic scatter from \cite{Marconi:2003fk}. \textit{Right}: $M_{BH}$ versus $M_{\rm bulge}$. The blue line and shaded region show the $M_{\rm bulge}-M_{BH}$ relation and scatter from \cite{Kormendy:2013ve}. }
\label{scalings}
\end{figure*}

In Figure~\ref{scalings} we show the position of RGG 118 relative to the $M_{\rm BH}-M_{\rm bulge}$ and $M_{\rm BH}-L_{\rm bulge}$ relations as defined by  \cite{Kormendy:2013ve} and \cite{2016ApJ...825....3L}. While the \cite{Kormendy:2013ve} relation is defined by elliptical/S0 galaxies with classical bulges, the \cite{2016ApJ...825....3L} relation includes late-types galaxies as well. The spiral galaxies from \cite{2016ApJ...825....3L} have BH masses ranging from $10^{6}-10^{8} M_{\odot}$. It is important to mention that the  \cite{2016ApJ...825....3L} sample is comprised of galaxies with BH masses measured dynamically (i.e., through megamasers). There are few low-mass galaxies for which comparisons between dynamical BH mass measurements and broad-line based measurements can be made. However, for NGC 4395, the broad-line mass from \cite{Reines:2013fj} is consistent with both the reverberation mapping mass \citep{2005ApJ...632..799P} and the recent gas dynamical measurement from \cite{2015ApJ...809..101D}. Additionally, the relationship between $R_{\rm BLR}$ and the 5100${\rm \AA}$ luminosity, on which the broad line mass measurements depend, has been shown to extend down to BHs of least $\sim7\times10^{5}$ $M_{\odot}$ \citep{2016ApJ...831....2B}. 

Our result is consistent with those of \cite{Greene:2008qy} (followed by \citealt{Jiang:2011vn, 2011ApJ...742...68J}) suggesting a break-down in scaling relations for low-mass ($M_{\rm BH}<10^{6}M_{\odot}$) BHs. Recent work by \cite{2015ApJ...813...82R} showed that nearby AGNs (including dwarf AGNs) fall systematically $\sim$ an order of magnitude below quiescent galaxies on the relation between total galaxy stellar mass and BH mass. This is potentially driven by a difference in host galaxy properties; they find a significant fraction of the AGN host galaxies are in spiral/disk galaxies. 

There are several possible explanations for why bulgeless/disk-dominated galaxies or those with pseudobulges do not correlate with BH mass in the same way as galaxies with classical bulges. \cite{Kormendy:2013ve} (see also \citealt{Greene:2008qy}; \citealt{Jiang:2011vn}) suggest that there are two different modes of BH growth: one in which a merger drives copious amounts of gas towards the center, growing the BH rapidly, and a second where BH growth is a local, stochastic process. The first mechanism would be relevant for BHs in bulge-dominated/elliptical galaxies, while BHs in disk-dominated galaxies would grow via the second mode. This is consistent with a picture in which the BHs in disk-dominated and/or pseudobulge galaxies are undermassive with respect to the scaling relations defined by relatively massive, classical bulge-dominated systems. 
\smallskip

In summary, we find that the light profile of RGG 118 is well described by an outer spiral disk, an inner S{\'e}rsic component with a stellar population older than $\sim1$ Gyr, and properties consistent with a pseudobulge, and a central point source. The properties of the central point source are consistent with originating from an AGN. We confirm that RGG 118 sits well below scaling relations between BH mass and bulge mass/luminosity, similar to other low-mass, disk-dominated systems.

\acknowledgements
 A.E.R. is grateful for the support of NASA through Hubble Fellowship grant HST-HF2-51347.001-A awarded by the Space Telescope Science Institute, which is operated by the Association of Universities for Research in Astronomy, Inc., for NASA, under contract NAS 5-26555.  Support for Program No. HST-GO-14187 was provided by NASA through a grant from the Space Telescope Science Institute, which is operated by the Association of Universities for Research in Astronomy, Inc., under NASA contract NAS5-26555. The authors thank Laura Ferrarese for helpful discussions. 

\software{AstroDrizzle (\footnotesize{drizzlepac.stsci.edu}),\normalsize{
GALFIT \citep{2002AJ....124..266P, 2010AJ....139.2097P},}
GALEV \citep{Kotulla:2009ul},
Starfit (\footnotesize{https://www.ssucet.org/$\sim$thamilton/research/starfit.html})}

\bibliographystyle{apj}

\clearpage

\end{document}